\documentclass[prb,reprint,superscriptaddress,amsmath,amssymb,nolongbibliography,aps]{revtex4-2}

\usepackage{graphicx}

\usepackage{color}

\usepackage{colortbl}
\usepackage{hyperref}
\usepackage{ragged2e}
\usepackage{enumitem}
\usepackage{cleveref}

\usepackage{tabularx}
\usepackage{siunitx}
\usepackage{multirow}
\usepackage{xcolor}

\usepackage{xr}
\newcommand{\squeezeup}{\vspace{-2.5mm}}
\usepackage{xpatch}
\usepackage[normalem]{ulem}
\xpatchcmd{\MaketitleBox}{\hrule\vskip12pt}{\vspace{-2\baselineskip}}{}{}% remove first horizontal rule (above abstract) + space
\xpatchcmd{\MaketitleBox}{\hrule}{}{}{}
\usepackage{array}
\newcolumntype{P}[1]{>{\centering\arraybackslash}p{#1}}
\usepackage{etoolbox}
\usepackage{comment}
\patchcmd{\abstract}{Abstract}{}{}{}

\usepackage{soul}

\renewcommand\thesection{\Roman{section}}

\begin{document}

\title{\textbf{High pressure melt dynamics in shock-compressed titanium}}
\author{Saransh Singh}
\email{saransh1@llnl.gov}
\affiliation{Lawrence Livermore National Laboratory, Livermore, CA, USA}

\author{Reetam Paul}
\affiliation{Lawrence Livermore National Laboratory, Livermore, CA, USA}

\author{Nikhil Rampal}
\affiliation{Lawrence Livermore National Laboratory, Livermore, CA, USA}

\author{Rhys J. Bunting}
\affiliation{Lawrence Livermore National Laboratory, Livermore, CA, USA}

\author{Sebastien Hamel}
\affiliation{Lawrence Livermore National Laboratory, Livermore, CA, USA}

\author{Nathan Pulver}
\affiliation{UCLA, Los Angeles, CA, USA}

\author{\\Christopher P. McGuire}
\affiliation{Lawrence Livermore National Laboratory, Livermore, CA, USA}

\author{Samantha M. Clarke}
\affiliation{Lawrence Livermore National Laboratory, Livermore, CA, USA}

\author{Amy L. Coleman}
\affiliation{Lawrence Livermore National Laboratory, Livermore, CA, USA}

\author{Cara Vennari}
\affiliation{Lawrence Livermore National Laboratory, Livermore, CA, USA}

\author{Trevor M. Hutchinson}
\affiliation{Lawrence Livermore National Laboratory, Livermore, CA, USA}

\author{\\Kimberly A. Pereira}
\affiliation{Department of Chemistry, University of Massachusetts Amherst, Amherst, Massachusetts 01003, USA}

\author{Bob Naglar}
\affiliation{Linac Coherent Light Source, SLAC National Accelerator Laboratory, Menlo Park, CA, USA}

\author{Dimitri Khaghani}
\affiliation{Linac Coherent Light Source, SLAC National Accelerator Laboratory, Menlo Park, CA, USA}

\author{Hae Ja Lee}
\affiliation{Linac Coherent Light Source, SLAC National Accelerator Laboratory, Menlo Park, CA, USA}

\author{Nicholas A. Czapla}
\affiliation{Linac Coherent Light Source, SLAC National Accelerator Laboratory, Menlo Park, CA, USA}

% \author{David McGonegle}
% \affiliation{AWE, }

\author{Travis Volz}
\affiliation{Lawrence Livermore National Laboratory, Livermore, CA, USA}

\author{Ian K. OCampo}
\affiliation{Lawrence Livermore National Laboratory, Livermore, CA, USA}

\author{James McNaney}
\affiliation{Lawrence Livermore National Laboratory, Livermore, CA, USA}

\author{Thomas E. Lockard}
\affiliation{Lawrence Livermore National Laboratory, Livermore, CA, USA}

\author{Jon H. Eggert}
\affiliation{Lawrence Livermore National Laboratory, Livermore, CA, USA}

\author{Amy Lazicki}
\affiliation{Lawrence Livermore National Laboratory, Livermore, CA, USA}

\author{Christopher E. Wehrenberg}
\affiliation{Lawrence Livermore National Laboratory, Livermore, CA, USA}

\author{Andrew Krygier}
\affiliation{Lawrence Livermore National Laboratory, Livermore, CA, USA}

\author{Raymond F. Smith}
\affiliation{Lawrence Livermore National Laboratory, Livermore, CA, USA}

% \author[add1]{Saransh Singh \corref{cor1}} \ead{saransh1@llnl.gov}
% \cortext[cor1]{Corresponding author}
% \author[add1]{Reetam Paul}
% \author[add1]{Nikhil Rampal}
% \author[add1]{Rhys J. Bunting}
% \author[add1]{Sebastien Hamel}
% \author[add2]{Nathan Pulver}
% \author[add1]{\\Christopher P. McGuire}
% \author[add1]{Samantha M. Clarke}
% \author[add1]{Amy L. Coleman}
% \author[add1]{Cara Vennari}
% \author[add1]{Trevor M. Hutchinson}
% %\author[add1]{\\Martin G. Gorman}
% \author[add3]{\\Kimberly A. Pereira}
% \author[add4]{Bob Naglar}
% \author[add4]{Dimitri Khaghani}
% \author[add4]{Hae Ja Lee}
% \author[add5]{David McGonegle}
% \author[add1]{Travis Volz}
% \author[add1]{\\Ian K. OCampo}
% \author[add1]{James McNaney}
% \author[add1]{Thomas Lockard}
% \author[add1]{Jon H. Eggert}
% \author[add1]{Andrew Krygier}
% \author[add1]{Raymond F. Smith}

% \address[add1]{Lawrence Livermore National Laboratory, Livermore, CA, USA}
% \address[add2]{UCLA, Los Angeles, CA, USA}
% \address[add3]{Department of Chemistry, University of Massachusetts Amherst, Amherst, Massachusetts 01003, USA}
% \address[add4]{Linac Coherent Light Source, SLAC National Accelerator Laboratory, Menlo Park, CA, USA}
% \address[add5]{Oxford Centre for High Energy Density Science, Department of Physics, Clarendon Laboratory, University of Oxford, Oxford, UK}

\date{\today}% It is always \today, today,

%\linenumbers
\begin{abstract}
\noindent
We study the high-pressure melting behavior of titanium using laser-driven shock compression with \textit{in situ} femtosecond x-ray diffraction and molecular-dynamics simulations based on a machine-learned interatomic potential. The MD simulations predict the solid-liquid coexistence on the Hugoniot in the $\sim$$111-124$ GPa range. Experimentally, we observe the first evidence of liquid at 86 GPa. We also observe pronounced microstructural  changes with pressure with strong grain refinement associated with the emergence of liquid, within the solid-liquid coexistence ($\sim$$110-126$ GPa). Above 126 GPa, we observe the persistence of residual levels of highly textured crystalline Ti to $\sim$$180$ GPa, well above the expected melt completion pressure. We discuss the accuracy that current laser-shock experimental platforms have at determining the melt onset and completion pressures. 
\end{abstract}

\maketitle

\section{Introduction}
\squeezeup
\noindent
The solid-to-liquid phase transition is among the most significant transformations occurring under shock compression, as it is accompanied by drastic changes in the mechanical and transport properties of materials. Accurate determination of melting under dynamic loading is therefore essential for reliable equation-of-state (EOS) models. Near-instantaneous shock compression is often the most direct means of accessing melting at extreme pressure ($P$) and temperature ($T$) conditions \cite{Wicks2024}. Among melt-related properties, the determination of solid--liquid coexistence along the Hugoniot provides a particularly stringent constraint on state-of-the-art multiphase semiempirical EOS models \cite{wu2023,singh2023}, motivating continued efforts to precisely identify both the onset and completion of melting under shock loading. More recently, shock-induced melting has attracted renewed interest due to observed deviations from classical nucleation-and-growth behavior: under rapid compression, phenomena such as orientation-dependent melt pressures \cite{renganathan2024,Renganathan2025} and the formation of disordered states at the shock front \cite{smith2026b} have been reported.

Historically, solid--liquid coexistence under rapid shock compression has been inferred from changes in the measured sound speed as a function of pressure in millimeter-thick samples \cite{nguyen2004}. Although this approach has provided valuable constraints, it relies on indirect signatures of phase transitions with no direct structural measurements. Interpretation of melt onset and completion is often complicated by significant scatter in the experimental data. Laser-driven shock compression coupled with in situ x-ray diffraction (XRD) offers a complementary diagnostic that directly probes the crystal structure of matter under dynamic loading. Recent advances in coupling ultrafast X-ray diffraction to dynamic compression platforms have enabled structural measurements under dynamic loading. This approach has become an increasingly powerful and widely adopted tool for investigating shock-induced melting \cite{briggs2019,coleman2022,singh2023,crepisson2025,Renganathan2025}.

In this study, we employ laser-shock compression with \textit{in situ} XRD to examine the melting transition in elemental titanium (Ti). Titanium and its alloys are of enduring technological and scientific importance due to their exceptional strength-to-weight ratio, corrosion resistance, and performance under extreme conditions, making them indispensable in aerospace, defense, and advanced manufacturing applications. Under compression, Ti exhibits a series of pressure- and temperature-induced structural transformations \cite{dewaele2015,akahama2020}, most notably involving transitions between the ductile $\alpha-$phase (hexagonal close-packed), the brittle $\omega-$phase (simple hexagonal), and the high-temperature $\beta-$phase (body-centered cubic). Elevated pressures and temperatures ultimately leads to melting from the $\beta-$phase. Accurate determination of melting conditions at high pressure ($P$), temperature ($T$), and strain rate therefore provides critical constraints for modeling extreme environments, such as those generated during high-velocity impacts. Consequently, the high-pressure phase diagram of titanium has been the subject of extensive theoretical \cite{kerley2003,greeff2001,cox2012} and experimental investigations using both static \cite{errandonea2001,stutzmann2015,dewaele2015,akahama2020} and dynamic compression techniques \cite{walsh1957,mcqueen1960,krupnikov1963,mcqueen1970,Marsh1980,al1981,trunin1999}.

Static compression studies have identified a series of low-$P$-$T$ structural phase transformations in titanium \cite{dewaele2015, akahama2020}. Under static compression, melt has been determined to lie within  $\sim$2400 $\rightarrow$ 2800 K at 100 GPa \cite{errandonea2001,stutzmann2015}, substantially lower than the temperatures predicted by several equation-of-state models, which place melting near 4300 K at the same pressure \cite{cox2012, kerley2003}. However, it remains unclear whether the static phase diagram of titanium adequately describes the states accessed under high–strain-rate dynamic compression, where time-dependent effects can significantly influence phase evolution, as observed in other metals \cite{Smith2013,coleman2019,gorman2019}.

In this work, we determine the phase stability of titanium at high $P$, $T$ and strain rate by combining  laser-shock compression with \emph{in situ} femtosecond x-ray diffraction (crystal structure, density, microstructural texture) and velocimetry (pressure determination). From 15$\rightarrow$66 GPa we record a phase evolution from $\alpha$ (hcp) $\rightarrow$ $\omega$ (simple cubic) $\rightarrow$ $\beta$ (bcc) phase transitions. Initially, in the $\beta-$phase, we observe a sharp texture in the XRD pattern consistent with large, highly-oriented grains. At the onset of melt, the  $\beta$-texture undergoes microstructural refinement leading to a more powder-like diffraction signature. The liquid phase fraction increases smoothly upon further compression of Ti up to near complete melt around 180 GPa.

% We also perform molecular dynamics (MD) simulations based on a machine-learned interatomic potential (MLIP), which predict a solid--liquid coexistence interval of 110--130~GPa, significantly narrower than the experimentally observed range of 86--179~GPa. At pressures well above the predicted completion of melting, we observe the persistence of residual crystalline $\beta$-Ti. We discuss the origins of the disagreement between theoretical predictions and experimental measurements, along with the inherent challenges in resolving solid--liquid coexistence using laser-shock compression combined with \textit{in situ} X-ray diffraction.

We also perform molecular dynamic (MD) simulations based on a machine-learned interatomic potential (MLIP) which predicts a solid-liquid coexistence between 110-130 GPa, a range significantly narrower than observed here experimentally (86–179 GPa). At high pressures, well above the expected completion of melt, we observe the persistence of residual crystalline grains of $\beta$-Ti. We discuss the possible reasons for the disagreement between theoretical and experimental coexistence and the inherent challenges in making solid-liquid coexistence measurements using laser-shock compression coupled with an \textit{in situ} X-ray diffraction diagnostic.

\section{Experimental Setup}
\begin{figure}[!t]
\begin{center}
\includegraphics[width=1\columnwidth]{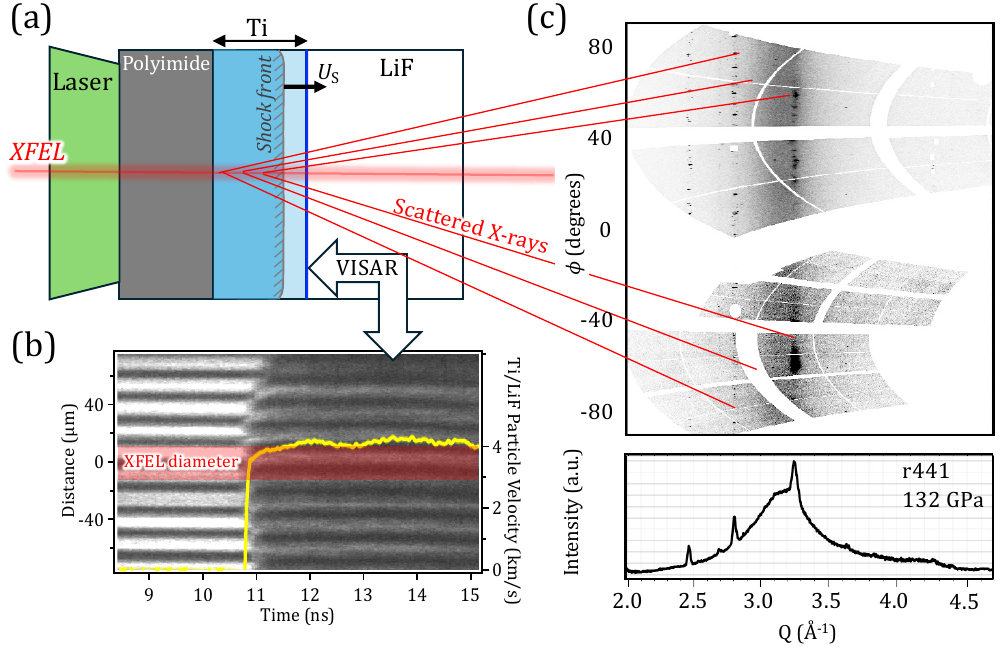}
\caption{\textbf{Experimental setup for high pressure structural measurements on shock compressed Ti. (a)} Target design consisting of  polyimide/Ti/LiF layers. \textbf{(b)} VISAR interferogram records the Ti/LiF interface velocity as a function of time (yellow trace), which allows for a determination of sample pressure during the x-ray probe time (see Supplemental Materials). \textbf{(c)} X-ray diffraction pattern provides data on crystal structure, density and microstructural texture within the compressed Ti. 
}
\label{fig:Ti_Setup}
\end{center}
\end{figure}

\squeezeup
\noindent
Laser-driven shock-compression experiments were performed at the Matter in Extreme Conditions (MEC) endstation of the Linac Coherent Light Source (LCLS) \cite{Nagler2015}. A schematic of the experimental setup and target design is shown in Fig. \ref{fig:Ti_Setup}(a). The target consisted of a 79 $\mu$m thick polyimide [C$_{22}$H$_{10}$N$_{2}$O$_{5}$] ablator layer, a 32 $\mu$m thick Ti polycrystalline foil, and a LiF window. A 0.2 $\mu$m thick Al layer was vapor deposited on the LiF to enhance reflectivity for velocimetry measurements at the Ti/LiF interface (Fig. \ref{fig:Ti_Setup}(a)). All target layers were bonded with a $\sim1\mu$m-thick layer of epoxy (Stycast 1266).

%The target assembly, as illustrated in Fig.  \ref{fig:Ti_Setup}A, consists of a 75 $\mu$m polyimide [C$_{22}$H$_{10}$N$_{2}$O$_{5}$] ablator layer adhered directly onto a $\sim$32 $\mu$m-thick polycrystalline Ti foil and a LiF window for velocimetry measurements.~A 0.1-$\mu$m coating of Al was applied to the LiF to enhance target reflectivity. All target layers were adhered together with a $\sim$1 $\mu$m thick glue layer.

The front surface of the polyimide ablator was positioned at the focal plane of four laser beams, which delivered a combined energy of up to 96~J at 527~nm in a 15~ns flat-top pulse focused to a $\sim300~\mu$m diameter spot (see Supplemental Materials). Laser energy absorbed by the polyimide ablator drives rapid ablation and expansion, generating a shock wave that propagates through the target assembly with a duration matched to the incident laser pulse. The pressure in the sample and the temporal steadiness of the compression wave were controlled by varying the total laser power and laser pulse shaping, respectively \cite{brown2017}. A line-imaging velocity interferometer system for any reflector (VISAR) \cite{celliers2023imaging} was used to determine the shock arrival at the Ti/LiF interface and to measure the Ti/LiF particle velocity history, $u_p(t)$ [Fig.~\ref{fig:Ti_Setup}(b)], which was used to constrain the sample pressure at the time of the x-ray probe (see Supplemental Materials).

\begin{figure*}[!t]
\begin{center}
\includegraphics[width=1\textwidth]{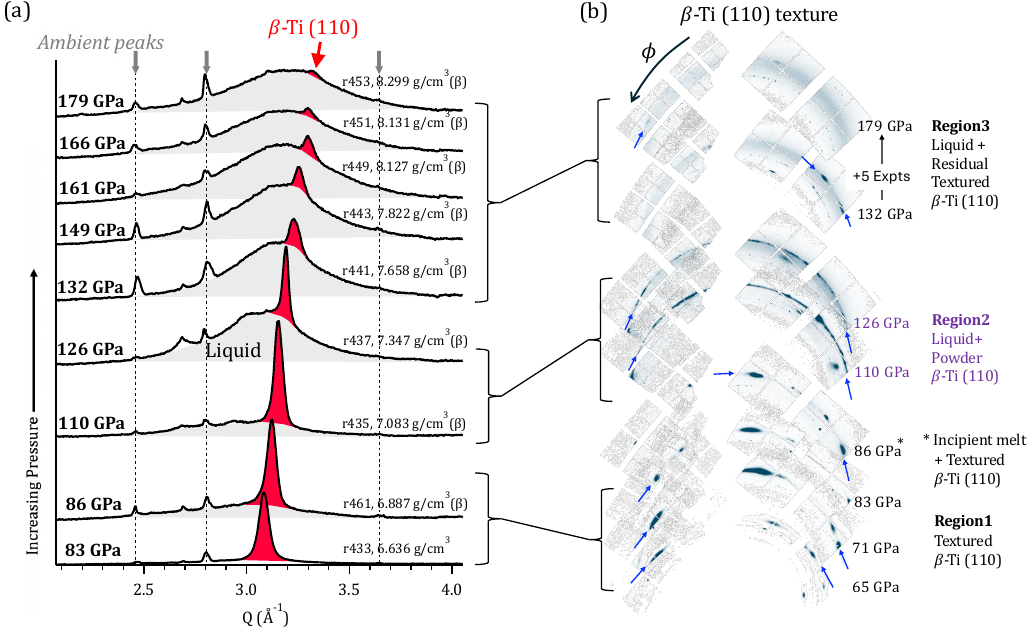}
%\captionsetup{width=0.5\linewidth}
\caption{\textbf{Ti x-ray diffraction profiles. (a)} Azimuthally-averaged diffraction pattern as a function of increasing shock pressure. The high pressure bcc $\beta-$phase is shaded in red, and the liquid diffraction signal is shaded in gray. Labels of experimental run number, pressure from VISAR, and $\beta-$phase density from XRD is labeled on each trace. The position of the uncompressed peaks (originating from regions of the sample ahead of the shock front) are highlighted by the vertical dashed lines. We observe the emergence of increased levels of liquid diffraction signal and a commensurate drop in scattering from the $\beta$-Ti (110) peak as a function of increasing shock pressure (see Fig.~\ref{fig:coexistence} for the $\beta$, liquid phase fractions as a function of pressure). \textbf{(b)} Select regions of x-ray diffraction pattern show the evolution of intensity distribution in the $(110)$ ring of the $\beta$-phase as a function of pressure. At pressures below melt (Region 1) the $(110)$ reflection is highly localized around the Debye-Scheerer cones, consistent with large orientated grains. Powder-like diffraction is observed within solid-liquid coexistence (Region 2), consistent with microstructure refinement. Finally, above the MLMD determined melt line (Region 3), a residual amount of highly textured $\beta$-Ti $(110)$ reflection is observed with diminishing intensity as a function of pressure.
}
\label{fig:Waterfall_texture}
\end{center}
\end{figure*}

The 50~fs output of the Linac Coherent Light Source (LCLS) X-ray free-electron laser (XFEL), operating at 10~keV, was incident on the target at normal incidence during shock transit through the titanium layer, focused to a 20~$\mu$m diameter spot centered on the laser drive. The relative timing and pointing between the X-ray probe and the laser-driven shock were controlled to within $<100$~ps and a few microns, respectively \cite{rastogi2022}. X-rays scattered from the compressed Ti were recorded in transmission geometry on multiple large-area ePix detectors \cite{Dragone2014}. The detector angular positions in $2\theta$ (diffraction angle) and $\phi$ (azimuthal angle around the Debye--Scherrer cones) were calibrated using diffraction patterns from an ambient-pressure CeO$_2$ standard \cite{tracy2019}. Representative diffraction data projected in linear momentum transfer $Q$ coordinates are shown in Fig.~\ref{fig:Ti_Setup}(c), where Bragg reflections appear as vertical features [Q = 4$\pi$sin($\theta$)/$\lambda$, where $\lambda$ is the x-ray wavelength and $\theta$ is the Bragg scattering angle]. Because the x-rays probe the sample during shock transit within the Ti layer, the resulting diffraction patterns are volume-integrated and contain contributions from both shocked and unshocked material.

\section{X-ray diffraction determination of structure}
\squeezeup
\noindent
In our X-ray diffraction measurements, we observe the expected phase sequence of $\alpha \rightarrow \omega \rightarrow \beta \rightarrow$ liquid with increasing pressure \cite{stutzmann2015,dewaele2015,errandonea2001,akahama2020}. Mixed-phase regions are identified for $\alpha+\omega$ at 15 and 22~GPa, and for $\omega+\beta$ at 36 and 49~GPa. Between 65 and 83~GPa, only the $\beta$ phase is observed (see Fig.~\ref{fig:Ti_Xray_profiles} for a summary of the full data set).

\begin{figure}[!t]
\begin{center}
\includegraphics[width=1\columnwidth]{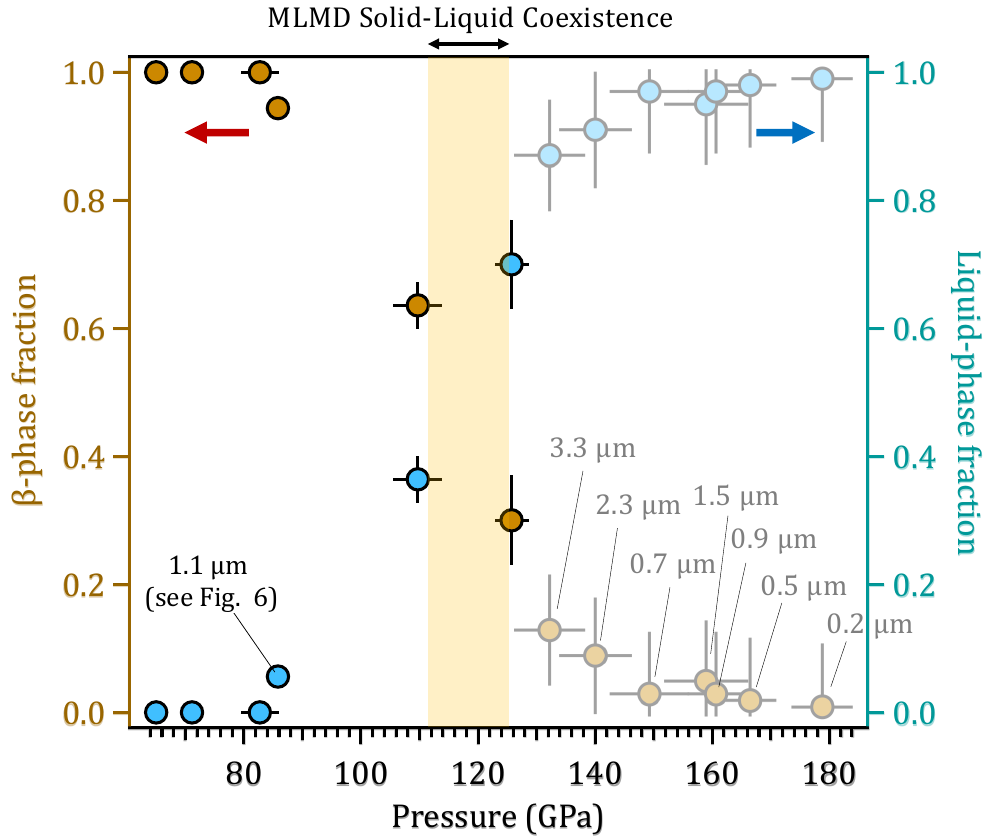}
%\captionsetup{width=0.5\linewidth}
\caption{\textbf{Phase fraction as a function of pressure.} $\beta$ and liquid phase fraction evolution as a function of pressure. Incipient melting is observed at 86 GPa (see Fig. \ref{fig:Ti_Xray_profiles_zoom}). The yellow band represents the the extent of the solid-liquid coexistence, under shock compression, as calculated by MLMD simulations. %\sout{The data from Region 2 in Fig. \ref{fig:Waterfall_texture}(b) is shown within a yellow band. The extent of the yellow band is based on assuming a linear pressure dependence of $\beta$ phase fraction with pressure within the solid-liquid coexistence. Based on this analysis we determine the extent of the solid-liquid coexistence along the Hugoniot to be from 100 GPa (incipient melt) to 130 GPa (complete melt). We note that there is a residual level of $\beta$-phase at higher pressure.}
For the high pressure shots we report on the estimated thickness of $\beta$-Ti present, based on a measure of the compressed sample volume during the x-ray probe period. The highest-pressure shots are shown as translucent to indicate reduced confidence in the phase-fraction estimates due to sharp texture and potentially insufficient grain-sampling statistics.
}
\label{fig:coexistence}
\end{center}
\end{figure}

Figure~\ref{fig:Waterfall_texture}(a) focuses on the high-pressure region probed in our experiment, showing azimuthally averaged diffraction profiles as a function of increasing pressure. From 86 to 179~GPa, a mixed $\beta+$liquid phase assemblage is observed. Over this pressure range, the fraction of the $\beta$ phase decreases with increasing pressure, such that at 179~GPa only a residual contribution from the strongest reflection, $\beta$(110), remains. This behavior is indicative of near-complete melting (see also Fig.~\ref{fig:coexistence}).

Fig.~\ref{fig:Waterfall_texture}(b) highlights the evolution of microstructural texture in the $\beta$-(110) reflection as a function of increasing shock pressure. The Ti microstructure demonstrates a pronounced change in its microstructure, as evidenced by the distribution of intensity around the Debye-Scherrer cones. Three distinct pressure regimes can be identified. Region 1 corresponds to the pre-melting regime (65–83 GPa), in which the diffraction intensity is azimuthally localized along the Debye–Scherrer cone, consistent with scattering from large, highly oriented $\beta$-Ti grains. Region 2 spans the solid–liquid coexistence regime (110–126 GPa) and is characterized by the emergence of broad, diffuse liquid scattering accompanied by a transition of the $\beta$-(110) reflection to a powder-like intensity distribution. At 86 GPa, a shot which straddles region 1 and 2, we observe incipient melt while the $\beta$-Ti remains strongly textured. At higher pressure, Region 3, exhibits a re-emergence of highly textured $\beta$-(110) diffraction with progressively diminishing intensity up to 179 GPa %\textcolor{blue}{\sout{As discussed below, we interpret the diffraction from Region 3 to be due to a cooler µm-thick layer of the Ti that is generated from pressure reverberation within the glue layer.}}

In Fig. \ref{fig:coexistence}, we present estimates of the $\beta$ and liquid phase fractions as a function of shock pressure (see Supplemental Materials for details).~After the first observation of liquid at 86 GPa, % pressures 110 and 126 GPa (Region 2 in Fig. \ref{fig:Waterfall_texture}(b)). 
%At higher pressures where residual textured $\beta$-phase diffraction is observed (Region 3 in Fig. \ref{fig:Waterfall_texture}(b)), the phase fraction assessment is based only the ratio of detected signal levels from both the liquid and the $\beta$-phase. Here 
we see a progressive drop in solid-phase fraction as function of increasing pressure, until at 179 GPa, only $\sim1$\% (0.2 µm integrated thickness) is $\beta$-Ti. %Based on the known shock velocity vs pressure relationship for Ti, for each shot we calculate the compressed Ti thickness at the x-ray probe time. Given this, and the estimate of phase fraction, we calculate the $\beta$-phase thickness as a function of pressure for our high pressure shots (labels in Fig. \ref{fig:coexistence}). 

Our experimental data suggests a larger pressure region of solid
+liquid (86$\rightarrow$179 GPa), than predicted from state-of-the-art theory (111$\rightarrow$124 GPa, yellow band in Fig. \ref{fig:coexistence}), which we describe in the following section.

%\textcolor{blue}{\sout{(I think we may not want to be so definitive at this point in the paper.) To determine the pressure extent of the solid-liquid coexistence along the Hugoniot we assume a linear dependence on solid fraction with pressure (solid line in Fig. \ref{fig:Waterfall_texture}). Based on this analysis we constrain the solid-liquid coexistence to be from 100-130 GPa.}}

\section{Machined Learned Molecular Dynamics (MLMD) Simulations}

\noindent
In the absence of a readily available temperature diagnostic in our experimental configuration, we do not measure the shock temperature. To calculate the the titanium melt line and Hugoniot we employ a multistage theoretical approach centered on density functional theory (DFT), machine learning, and molecular dynamics. This theoretical approach is described in detail in Supplemental Materials Section \ref{sec:simulations_SM}. Briefly, a DFT-generated dataset was constructed using ab initio molecular dynamics along multiple isochores spanning pressures from $\sim$100 to 180 GPa and temperatures between 2000 and 4000 K. These calculations employed the Perdew-Burke-Ernzerhof (PBE) \cite{PerdewBurkeErnzerhof1996} functional within the generalized gradient approximation as implemented in in the Vienna \emph{ab initio} simulation package (VASP) \cite{Kresse1996}, and were designed to sample diverse atomic environments. The resulting dataset of energies and forces were then used to train a machine-learned interatomic potential (MLIP) through the Allegro framework, which showed excellent agreement with the underlying DFT values.

\begin{figure}[!t]
\begin{center}
\includegraphics[width=1\columnwidth]{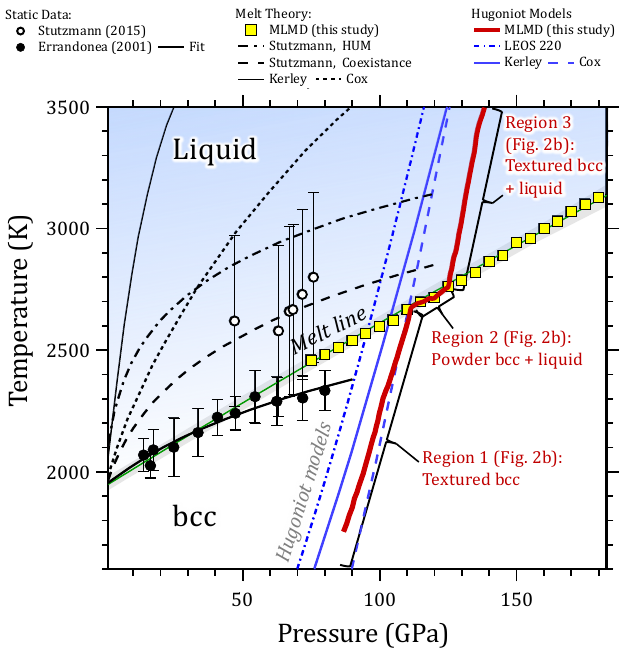}
%\captionsetup{width=0.5\linewidth}
\caption{\textbf{$P$-$T$ melt in titanium.} The high pressure onset of melt has been constrained under static compression by Stutzmann \emph{et al.} \cite{stutzmann2015} (open circles) and Errandonea \emph{et al.} \cite{errandonea2001} (filled circles). The melt line as determined by our MLMD simulations are defined by the yellow squares. A fit to these data (with shaded uncertainties) is represented by the green curve. The calculated Hugoniot states are shown by the solid red line. The Hugoniot states from other EOS models shown are Kerley \cite{kerley2003}, Cox \cite{cox2012} and Livermore-EOS (LEOS) model \#220.~The three $P$-$T$ regions with differing $\beta-$phase microstructure detailed in Fig.~\ref{fig:Waterfall_texture}(b) are labeled here. %A schematic of the $\beta-$phase microstructure is shown for the three regions. Below the solid-liquid coexistence (Region 1), the $\beta-$phase has large, highly oriented grains. In the solid-liquid coexistence region (Region 2), we observe microstructure refinement. Since there is a sudden and drastic change in the microstructure, we picture this region as a collection of $\beta-$phase surrounded by the liquid phase facilitating grain reorientation. Above the coexistence, we observe residual grains of the $\beta-$phase. Since we rule out all experimental artifacts that could give rise to the persistence of these grains well-above the predicted melt line, we attribute this effect to orientation-dependent melt behavior in the near-instantaneous extreme strain-rate shock compression. 
}
\label{fig:Ti_Phase_Map}
\squeezeup
\squeezeup
\squeezeup
\end{center}
\end{figure}

Using this MLIP, we performed molecular dynamics (MD) simulations across a broad pressure–temperature range to map the titanium equation of state (EOS). The MLIP enabled longer sampling times and larger system sizes, allowing us to simulate melting with two-phase (solid–liquid) configurations and to carry out extended NPH (constant number, pressure, and enthalpy) runs to determine solid–liquid coexistence conditions. From these calculations we obtained the titanium melting curve over the explored pressure range (yellow squares in Fig.~\ref{fig:Ti_Phase_Map}). We fit a Simon-Glatzel curve to the melt data determined from the molecular dynamics simulation. The melt temperature as a function of pressure is given by
\begin{equation}
    T = 1941\times\left(\frac{P}{224.2\pm 34.4} + 1\right)^{0.8034\pm0.1}
    \label{eq:Tmelt} ,
\end{equation}
with pressure in GPa and temperature in Kelvin. Our calculated melt data aligns with the low pressure melt points from the static compression study of Errandonea \emph{et al.} \cite{errandonea2001}.

The EOS dataset was also used to calculate the Hugoniot %: the Hugoniot function was derived directly from the Rankine–Hugoniot relation, and the principal Hugoniot was identified as the locus of points where this function vanishes 
(red curve in Fig. \ref{fig:Ti_Phase_Map}). %Furthermore, isentropes were obtained by integrating thermodynamic relations from the EOS data, allowing comparison with the melt line.
The combined EOS, Hugoniot, and melting results revealed that the Hugoniot intersects the melt line between $\sim$111 and 124 GPa (yellow band in Fig. \ref{fig:coexistence}), consistent with experimental observations over liquid + powder $\beta$-Ti (region 2 in Fig. \ref{fig:Waterfall_texture}), but much less that the overall experimental observation of solid-liquid coexistence. %\sout{This theoretical framework thus provided both a predictive tool for interpreting shock-compression data and a means to validate experimental determinations of titanium’s high-pressure melting behavior.}

\section{Discussion\label{sec:discussion}}
\squeezeup
\noindent
We observe a significant discrepancy between the pressure range of solid–liquid coexistence predicted by the machine-learned molecular dynamics simulations and that inferred from our laser-driven shock-compression experiments. While the simulations predict coexistence over a relatively narrow pressure interval, the experimental measurements indicate a substantially broader range, extending both to lower pressures at the onset of melting and to higher pressures beyond the predicted completion of melt. In the following sections, we examine the potential origins of this discrepancy by evaluating limitations associated with the experimental platform and diagnostics, as well as assumptions inherent to the simulation methodology. Specifically, Section~\ref{sec:experimental} discusses possible experimental sources to the discrepancy while Section~\ref{sec:simulation} discusses possible limitations in the theoretical determination of solid–liquid coexistence.

\begin{figure}[!t]
\begin{center}
\includegraphics[width=1\columnwidth]{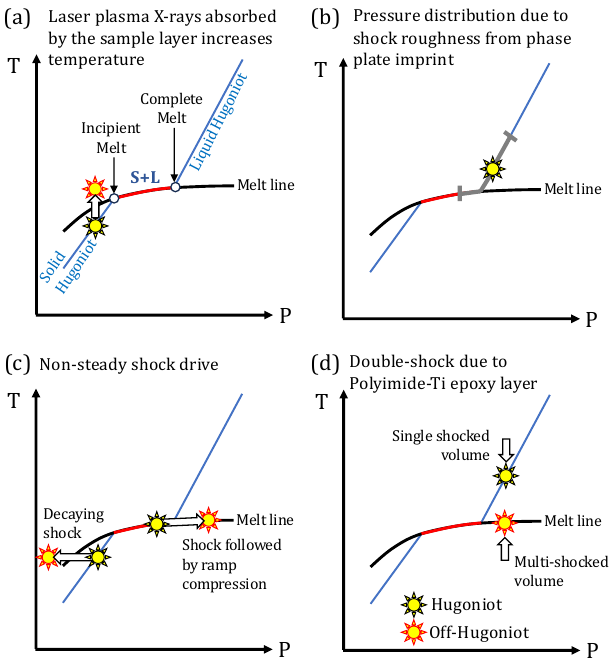}
\end{center}
\caption{\textbf{Representation of potential contributors to $P$-$T$ gradients when using laser-shock experiments to determine solid-liquid coexistence.} A $P$-$T$ phase map illustrating expected equilibrium behavior as the Hugoniot intersects with the melt line. Here, the red curve represents the region of solid (S) - liquid (L) coexistence. For each panel the black star represents an on-Hugoniot state, whereas the red star represents a portion of the sample which is off-Hugoniot. Potential experimental contributors include:~\textbf{(a)} Preheating of sample due to laser-plasma x-rays, ~\textbf{(b)} Phase plate imprint \cite{gorman2022}, ~\textbf{(c)} Temporally non-steady shock states, and \textbf{(d)} A $\sim$$\mu$m-thick epoxy layer resulting in a low-$T$ double-shock region \cite{coleman2022}. See text for details.
}
\label{fig:Ti_Equilibrium_Melting}
\squeezeup
\end{figure}

\subsection{Experimental uncertainties in constraining the solid-liquid coexistence \label{sec:experimental}} 
\noindent
The accuracy with which the pressure extent of solid–liquid coexistence can be determined in laser-driven shock experiments depends on three primary factors: (1) the sensitivity of the diagnostic to small volumes of liquid (incipient melting) and residual solid (completion of melting), (2) the precision with which pressure can be determined, and (3) the ability to generate spatially and temporally uniform compression states in $P$-$T$-$\rho$ (Fig.~\ref{fig:Ti_Equilibrium_Melting}). The first factor is governed by the x-ray diffraction diagnostic, while the latter two are specific to the laser-shock platform and target design. In general, limitations in diagnostic sensitivity tend to reduce the inferred pressure range of coexistence, whereas uncertainties in pressure determination and compression uniformity tend to broaden it. In this section, we describe additional experiments and simulations performed to quantify these uncertainties and assess their contributions to the experimentally observed solid–liquid coexistence range. 

\subsubsection*{Minimum detectable liquid volume}
\noindent Unlike the solid diffraction which is highly localized, liquid diffraction is diffuse in nature. This makes detection of small volumes of liquid difficult with an x-ray diffraction based diagnostic. Here, we estimate the minimum thickness of liquid Ti detectable using the ePix detectors at LCLS operating at 10 keV XFEL energy. We shock-compressed a 1-$\mu$m-thick Ti foil to pressures within the coexistence region, and measured the x-ray diffraction signal from the solid-liquid mixture. We determined the thickness of liquid Ti to be $\sim 0.85 \ \mu$m for this experiment with a signal-to-noise (S/N) ratio of $\sim10$ (see Supplemental Materials Section \ref{sec:detectability} and Fig.~\ref{fig:Ti_liquid_sensitivity}).

In an experiment using thicker Ti foils, the liquid diffraction signal is attenuated by self-absorption by the sample. This is expected to deteriorate the minimum detectable thickness of the liquid. To estimate the minimum detectability in the presence of extra absorption, we make the assumption that the self-absorption only reduced the signal and not the noise. The minimum detectable limit was estimated from the the previously described experiment by including self-absorption from the thicker Ti layer. We determine that the self absorption of the $32~\mu$m thick Ti sample is sufficient to reduce the S/N from $10$ measured in these experiments to just below $2$ (see the Supplemental Materials section~\ref{sec:detectability}). Therefore, based on our X-ray diffraction measurements on a $1 \ \mu$m thick foil, we determine the minimum detectable effective thickness of liquid Ti in a $\sim$32-$\mu$m-thick Ti sample to be $\sim$0.85-$\mu$m. The readers are referred to Supplemental Material Section \ref{sec:detectability} for a detailed review of the liquid detectability limit. 

\begin{figure}[!t]
\begin{center}
\includegraphics[width=1\columnwidth]{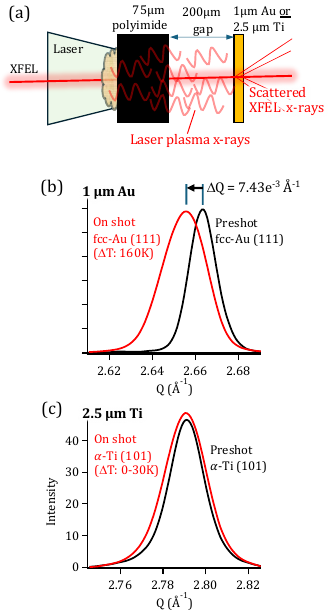}
\end{center}
\caption{\textbf{Experimental determination of heating from laser plasma x-rays. (a)} The experimental setup employs a 200-µm gap to temporally decouple heating (latice expansion) from laser plasma x-rays from the subsequent compression wave.~A laser intensity of 1.2$\times$10$^{13}$ W/cm$^2$ was used (115 J over 10-ns), with an XFEL probe time of 10-ns \cite{smith2026}. \textbf{(b)} Lattice expansion of a 1-µm-thick Au foil was determined to be equivalent to 160-K of heating. \textbf{(c)} Similar measurements on a 2.5-µm-thick Ti foil yielded no measurable heating effects.
}
\label{fig:Xray_heating}
\end{figure}

\begin{figure}[!t]
\begin{center}
\includegraphics[width=1\columnwidth]{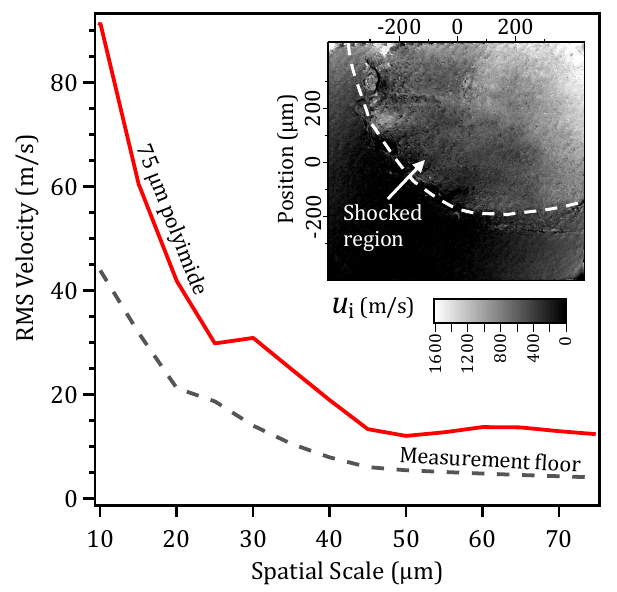}
\end{center}
\caption{\textbf{Estimate of shock roughness emerging from polyimide ablator layer due to phase plate imprint. Adapted from Ref. \cite{gorman2022}.} 
The inset figure represents a two dimensional velocity map of shock roughness using a comparable $527$-nm laser-shock drive on the Janus laser facility.~The main figure is the determined root-mean-squared (RMS) particle velocity roughness emerging from the 75-$\mu$m-thick polyimide ablator. For a 20-$\mu$m diameter XFEL spot, as used in our study, we estimate an additional $2$-GPa increase to the pressure distribution applied to the Ti layer. %We treat this value as a upper bound due to the spatially overlapping $2$ laser beams at the MEC endstation compared to the Gorman \emph{et al.} \cite{gorman2022} study, as well as the smaller characteristic laser speckle size at MEC \cite{gorman2022}. 
The pressure distribution can be reduced further if a larger diameter XFEL spot is used \cite{gorman2022}.
}
\label{fig:Ti_roughness}
\end{figure}

%\textcolor{red}{(need to check this value with Saransh, as well as the following paragraph which seems somewhat of a repeat. Was a certain level of TDS assumed for these liquid sensitivity determinations?)}

\subsubsection*{Laser plasma heating (Figs.~ \ref{fig:Ti_Equilibrium_Melting}(a), \ref{fig:Xray_heating})\label{sec:plasma_heating}} 
\noindent
High-pressure states in our experiments are generated by laser ablation of 75-µm-thick polyimide. The ablation plasma generated in these experiments produces x-rays in the sub- to few-keV range. In our experiments, these x-rays can potentially be transmitted through the polyimide layer and be absorbed by the Ti \cite{smith2026}. This would result in portions of the Ti being hotter than the Hugoniot, allowing melt to be accessed at pressures below the Hugoniot melt pressure (see Fig. \ref{fig:Ti_Equilibrium_Melting}(a)).

%pre-heat the sample before the pressure wave from the ablation reaches it. The consequence is that the sample $P$-$T$-$\rho$ state does not lie on the principle Hugoniot as shown schematically in Fig.~\ref{fig:Ti_Equilibrium_Melting}\textbf{(a)}. 
To quantify the level of pre-heating due to the laser plasma,~we conducted additional experiments at the MEC endstation of LCLS. A schematic of the experiment is shown in Fig. \ref{fig:Xray_heating}(a). The target consisted of 75-µm-thick polyimide and a 1-µm Au or 2.5-µm Ti sample layer. There was a 200-µm vacuum gap between the polyimide and the sample layer to isolate the contribution of laser plasma heating from hydrodynamic heating directly from the shock wave.~Laser-plasma heating expands the lattice of the sample, the extent of which is directed measured through x-ray diffraction (see Ref. \cite{smith2026} for details). Lattice expansion is related to sample temperature through the previously determined lattice expansion coefficients for Au and Ti. %to determine the extent of lattice expansion due to heating from laser plasma x-rays. We correlate the expanded volume of the lattice to a temperature using the known ambient thermal expansion coefficients. 
To maximize the heating effect, we employed a laser intensity of 1.2$\times$10$^{13}$ W/cm$^2$, or twice what was used in our primary study on Ti.

Figs. \ref{fig:Xray_heating}(b) and (c) summarize the results of this study. Here we plot the primary diffraction peak for both Au and Ti (ambient and heated) as a function of Q. Under the laser conditions described above, we observed consistency with a $\sim160$-K temperature rise in a 1-µm-thick Au foil and no heating from a 2.5-µm-thick Ti foil within our measurement uncertainty ($\sim30$K). This measurement uncertainty arises primarily from the small shot-to-shot shifts in the XFEL spectra in the SASE mode.

To extend the validity of these measurements to an experimental configuration in which the effective ``gap'' between the ablator and the sample approaches zero, we scale the heating by the ratio of the corresponding view factors for the two gap thicknesses (see Supplemental Materials Fig.~\ref{fig:view_factor}). Because we do not observe any heating-induced shift of the peak position for the Ti sample (i.e., no measurable heating), the view-factor correction provides an upper bound on the temperature increase of $\sim80$--$120~\mathrm{K}$, based on our measurement uncertainty ($\sim30~\mathrm{K}$). The actual temperature increase is likely smaller. Therefore, given the comparable absorption of $1~\mu\mathrm{m}$ Au and $32~\mu\mathrm{m}$ Ti for the expected few-keV laser-plasma x rays, together with the $\sim4\times$ higher specific heat of Ti relative to Au, we conclude that heating of our $32~\mu\mathrm{m}$-thick Ti samples by laser-plasma x rays is negligible in the primary Ti study.

\subsubsection*{Shock front roughness (Figs. \ref{fig:Ti_Equilibrium_Melting}(b), \ref{fig:Ti_roughness})}
\noindent
The use of continuous phase plates (CPPs) is common practice in laser experiments at XFELs in order to generate a spatially averaged uniform laser intensity profile with shot-to-shot consistency. The laser focal spot intensity profile is characterized by a high-contrast, high-frequency laser speckle. Without sufficient smoothing, these laser non-uniformities can translate to a significant pressure distribution within the sample layer probed by the XFEL. In an effort to quantify this effect, we refer to the study by Gorman \emph{et al.} \cite{gorman2022} as summarized in Fig. \ref{fig:Ti_roughness}. Here, the shock roughness emerging from a 75-µm-thick polyimide ablator, generated from a single beam of the Janus laser, is measured using two-dimensional velocimetry techniques \cite{smith2013b}. Based on these data, we estimate that, for our experimental geometry, this will result in a 2 GPa increased pressure distribution within our Ti sample. We consider this to be an upper limit as MEC employs a two laser beam overlap geometry and has a smaller characteristic laser speckle size compared to the Gorman \emph{et al.} \cite{gorman2022} study. Both of these attributes will serve to reduce the shock roughness further. We note that this pressure distibution is expected to be larger for experiments which use thinner polyimide layers \cite{gorman2022}. 

\begin{figure}[!t]
\begin{center}
\includegraphics[width=1\columnwidth]{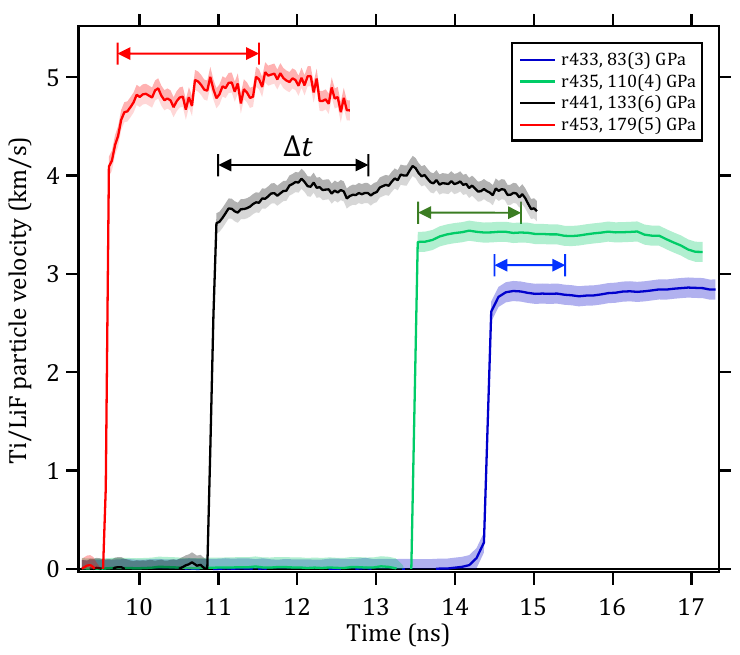}
\end{center}
\caption{\textbf{Representative Ti/LiF interface velocity vs time traces.} For each shot the shaded region represents the uncertainty in velocity determination (estimated here to be 5\% of a fringe shift). The temporal portion of the velocity trace used for each shot to determine pressure, $\Delta$t, is governed by the x-ray probe time relative to shock arrival at the Ti/LiF interface, and the pressure-dependent shock velocity.~As described in Supplementary Materials, the temporal steadiness (velocity distribution) over $\Delta$t is used to determine the central pressure and pressure distribution in the Ti sample at the x-ray probe time.
}
\label{fig:VISAR_lineouts}
\end{figure}

\bigskip
\subsubsection*{Non-steady laser drive (Fig. \ref{fig:Ti_Equilibrium_Melting}(c)),\\ and pressure determination accuracy (Fig. \ref{fig:VISAR_lineouts})}
%For (3) there are a number of potential contributors as illustrated in Fig. \ref{fig:Ti_Equilibrium_Melting}(a)$\rightarrow$(d): (a) X-ray heating from the laser plasma could generate higher-$T$ off-Hugoniot states which access the melt line at lower pressures than the equilibrium Hugoniot; (b) The impedance mismatch due of the epoxy at the polyimide-Ti interface results in a reverberation as the shock wave passes through. This results in a cooler double-shocked region in the Ti sample (see Ref. \cite{coleman2022}). (c) 
\noindent
In laser-shock experiments, temporal unsteadiness of pressure states after the shock front will result in the generation of off-Hugoniot states due to either: (i) isentropic pressure-release
from an initial shock state in the case of an unsupported shock, or (ii) shock + ramp–compressed states in the case of a growing shock. For (i) the $P$-$\rho$ states produced are less dense than the Hugoniot and could cross the melt line at lower pressure, and for (ii) the $P$-$\rho$ states produced are denser than the Hugoniot and could remain on the melt line for pressures higher that the the equilibrium solid-liquid coexistence (Fig. \ref{fig:Ti_Equilibrium_Melting}(c)). We note, as shown in Fig. \ref{fig:Ti_Laser_Pulse_Shaping}, that the majority of the shots in our experiments are best characterized by (ii).

In Fig. \ref{fig:VISAR_lineouts} we show representative Ti/LiF velocity profiles over the 83-179 GPa pressure range. The determined average pressure uncertainty is $\pm$3-6 GPa. This represents a limit of the accuracy, in our current study, with which we can determine the pressure onset or completion of melt. 

\begin{figure}[!t]
\begin{center}
\includegraphics[width=1\columnwidth]{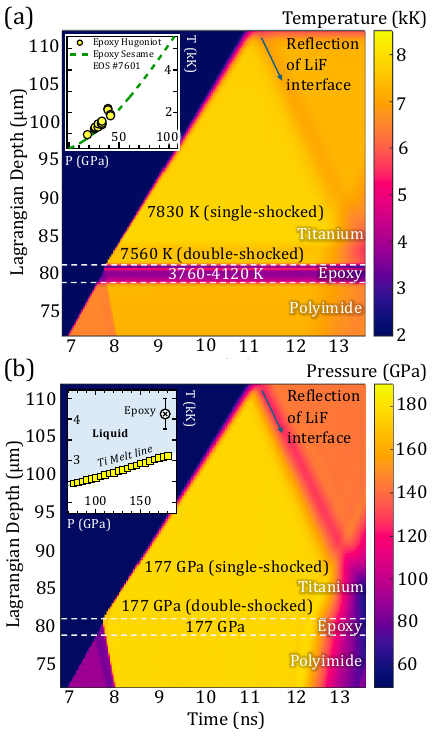}
%\captionsetup{width=0.5\linewidth}
\caption{\textbf{Calculation of high-$P$, low-$T$ layer in Ti sample.} \textbf{(a)} and \textbf{(b)} show the calculated evolution of temperature and pressure
respectively in the Ti target assembly from a HYADES hydrocode \cite{Larsen1994} simulation. Here we assume an epoxy thickness of 2-µm. Shock entry into the epoxy is followed by rapid wave reverberations, which ultimately produces a cooler few-µm thick double shocked region in the Ti layer. For bulk pressures just above full melt, crystalline $\beta$-Ti diffraction can persist from this cooler region. However, for the highest pressure shots in this study this is not considered a likely mechanism for the observed residual $\beta$-Ti diffraction.% even when the bulk material is fully melted.   
}
\label{fig:Ti_Hyades}
\end{center}
\squeezeup
\squeezeup
\end{figure}

\subsubsection*{Pressure reverberation in the epoxy layer (Figs.~\ref{fig:Ti_Equilibrium_Melting}(d), \ref{fig:Ti_Hyades})}
\squeezeup
\noindent
We ran one-dimensional HYADES hydrocode \cite{Larsen1994} simulations of our experimental geometry (Fig. \ref{fig:Ti_Setup}(a)), with the laser power chosen to generate a pressure of 177 GPa within the Ti sample. This pressure is close to the maximum achieved in our study and corresponds to conditions under which residual $\beta$-Ti was still observed (Fig. \ref{fig:Waterfall_texture}). The goal of these simulations was to determine whether temperature gradients within the sample, arising from the epoxy layer bonding the Ti foil to the polyimide ablator, could explain the presence of $\beta$-Ti in the XRD data.

HYADES calculates
the hydrodynamic flow of temperature and pressure waves through the target assembly in time and space.~The inputs to the hydrocode are the thicknesses of each of the layers within the target, including an EOS description of each of the materials within the target, and laser intensity as a function of time. The simulations used a single phase EOS for Ti (Livermore EOS \#220, see Fig. \ref{fig:Ti_Phase_Map}). Above melt this EOS is expected to produce Hugoniot temperatures which are $\sim$1000-K hotter than the MLMD multi-phase Hugoniot. The EOS used for epoxy was Sesame Table \#7601.~The inset to Fig. \ref{fig:Ti_Hyades}(a) shows the $P$-$T$ Hugoniot for epoxy from the Sesame EOS plotted against previously reported shock data (circle symbols) \cite{bordzilovskii2016}. There was no direct preshot measurement of the epoxy layer thickness. Therefore, based on previous measurements of similar target types, we assumed a 2-µm-thick epoxy layer for these simulations, which we consider an upper limit. As the thermal conductivities of Ti and epoxy at high pressure and temperature are not well constrained, we use their reported ambient values.~The pressure and temperature distributions throughout the sample layers are plotted as a function of time and Lagrangian depth in Fig. \ref{fig:Ti_Hyades}.

% \begin{comment}

% \end{comment}

For Hugoniot pressures within the stability field of the liquid (above complete melt) there are two mechanisms by which the presence of an epoxy layer could serve to cool a region of the Ti sample sufficiently to produce $P$-$T$ states which maintain residual levels of $\beta$-Ti: (i) As detailed in Ref.~\cite{coleman2022}, due to pressure reverberations within the epoxy layer at the polyimide-Ti interface, a thin layer of Ti experiences a multi-shock (lower-$T$ off-Hugoniot) compression state, (ii) the temperature of the epoxy layer is less than than the multi-shocked Ti layer. Heat flow from Ti to epoxy will further serve to cool this multi-shocked region.

From the HYADES simulations in Fig. \ref{fig:Ti_Hyades}, we estimate that the multi-shocked Ti layer is $\sim$270-K lower temperature compared to the single-shocked bulk Ti. The thickness of this layer depends on the pressure-dependent shock speed in the epoxy layer, which sets the reverberation round trip time, and can be comparable to the thickness of the epoxy layer itself. %We note, however, that there was no direct preshot measurement of this glue layer thickness, and reverberations within the epoxy layer would not be manifest in velocity measurements at the Ti/LiF interface. 
We estimate the bulk temperature (single-shocked region) at 177 GPa is $\sim$2400-K above the calculated melt line in Fig. \ref{fig:Ti_Phase_Map}, and therefore the formation of a multi-shocked layer alone cannot explain the persistence of residual $\beta$-Ti at this pressure. We also considered potential heat flow from Ti into the epoxy layer. %, for a range of possible thermal conductivities. 
The inset to Fig. \ref{fig:Ti_Hyades}(b) shows that the calculated epoxy temperature at 177 GPa is $\sim$1000-K above our estimated Ti melt line, and so any potential heat flow would not cool the Ti sufficiently. %However, under the shock loading conditions needed to achieve a 177 GPa shock in Ti the calculated temperature in the epoxy layer is 3783 K, which is $\sim$700-K hotter than the calculated melt line. 
%We therefore conclude that the presence of an epoxy layer in intimate contact with the Ti foil, does not result in $P$-$T$ states which explain the observed residual levels of $\beta$-Ti at high pressure. 

However, for lower $P$-$T$ Hugoniot states just above full-melt, the epoxy layer is expected to sufficiently cool the Ti to produce residual and detectable levels of $\beta$-Ti. This  confounds the experimental determination of completion of melt. While the high-$P$ conditions shown in Fig. \ref{fig:Ti_Hyades} show that $T$-gradients are not sufficient to explain residual levels of $\beta$-Ti at the highest pressures of our study, we recognize that $\beta$-Ti could be observed at residual levels \underline{if} the calculated temperature of the epoxy layer is cooler than the EOS models suggest and \underline{if} the heat flow from the hot-Ti into the epoxy is sufficiently high.

\subsection{Simulation limitations in constraining the solid-liquid coexistence\label{sec:simulation}} 
\noindent
The simulations employed in this work have several limitations that may limit direct comparability with coexistence pressures inferred from shock-compression experiments. First, the molecular dynamics simulations describe solid–liquid coexistence under equilibrium conditions in defect-free single crystals. As such, they do not capture kinetic effects or the influence of microstructural defects such as grain boundaries, dislocations, or vacancies. Second, the applied thermostat maintains thermal equilibrium across all the atoms. This contrasts with shock-compression experiments, where phonon transport can limit heat redistribution within the material and lead to spatially heterogeneous temperature field.

Additional uncertainties arise from approximations inherent to quantum chemical calculations. The exchange–correlation functional used in the underlying density functional theory calculations introduces errors that propagate through the machine-learned interatomic potential. Furthermore, residual training errors in the MLIP may affect its ability to accurately reproduce forces and energies across the full range of atomic environments sampled during melting.

% \textcolor{blue}{
% [Rhys and Nikhil to elaborate]
% \begin{itemize}
%     \item DFT functional is an approximation.
%     \item MAE in DFT energies and forces (very small for the Ti potential)
%     \item Simulations are for thermodynamic equilibrium while experiments could be non-equilibrium with kinetic effects.
%     \item Real Ti sample is polycrystalline and has defects such as vacancies, dislocations and grain boundaries, while the MD simulations assume a perfect single crystal. This effect will be important for determining the incipient melting conditions.
% \end{itemize}
% }

\begin{figure}[!t]
\begin{center}
\includegraphics[width=1\columnwidth]{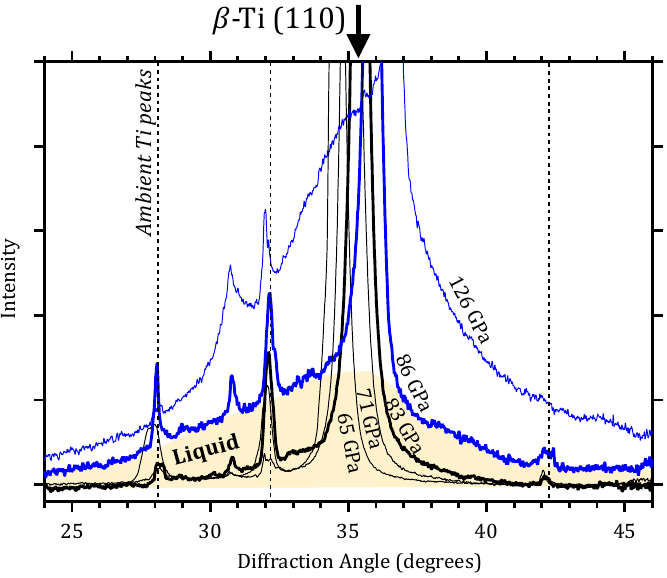}
%\captionsetup{width=0.5\linewidth}
\caption{\label{fig:Ti_Xray_profiles_zoom} \textbf{Transition from compressed bcc-only observation to bcc with liquid.} The onset of melt is determined by the first observation of angularly broad diffuse scattering. In our data that occurs at an estimated shock pressure of 86 GPa, with liquid contribution highlighted with yellow shading. We note that for pressures up to 83 GPa the diffuse signal around the $\beta$-Ti (110) peak we attribute to thermal diffuse scattering (TDS) \cite{wark2025}.
\label{fig:TDS}
}
\end{center}
\end{figure}

\subsubsection*{Thermal diffuse scattering (Fig.~\ref{fig:TDS})}
\squeezeup
\noindent
For a wide range of materials, the melt transition occurs at very high temperatures ($1000$s of K) under shock compression. At these conditions, the inelastic scattering due to the thermal vibrations of individual atoms can be a significant fraction of the total scattered intensity. The measured signal due to inelastically scattered photons, commonly referred to as thermal diffuse scattering (TDS), results in diffuse intensity between the main peaks \cite{warren1990,wark2025,heighway2025}. Therefore, the first observation of diffuse scattering as a result of TDS can be misinterpreted as evidence for incipient melting. A correct determination of incipient melting should take TDS into account. 

Fig.~\ref{fig:TDS} shows the azimuthally-averaged signal for shots with increasing shock pressure. Using the thermal diffuse model for powder BCC crystal described in Ref.~\cite{warren1990}, we could attribute the weak diffuse signal for all shots up to 83 GPa to thermal diffuse scattering. However, at 86 GPa, there is a sudden increase in the amount of diffuse signal which could not be attributed solely to TDS. Therefore, we determined the incipient melt pressure in our experiments to be 86 GPa. We note that the TDS model used in this work did not include the texture of $\beta-$Ti. However, as shown in Ref.~\cite{heighway2025}, the TDS signal is largely insensitive to the texture of the polycrystal, and can be safely ignored.

\section{Conclusions}
\noindent
We have investigated the melting behavior of titanium under laser-driven shock compression using \textit{in situ} femtosecond x-ray diffraction combined with velocimetry, providing direct structural constraints on phase evolution to nearly 180~GPa. Our measurements resolve the expected $\alpha \rightarrow \omega \rightarrow \beta$ phase sequence at intermediate pressures and reveal pronounced microstructural evolution within the high-pressure $\beta$ phase as the Hugoniot approaches and traverses melting conditions. In particular, the onset of melting is marked not only by the emergence of diffuse liquid scattering but also by a rapid transformation of the $\beta$-Ti microstructure from highly textured, large grains to a refined, powder-like distribution.

Quantitative phase-fraction analysis shows that solid--liquid coexistence persists over a broad pressure interval ($\sim$86--179~GPa), substantially wider than the coexistence region ($\sim$110--130~GPa) predicted by machine-learned molecular dynamics simulations trained on density functional theory data. At pressures well above the predicted completion of melting, weak but highly textured $\beta$-Ti diffraction persists, indicating that a small fraction of crystalline material survives to conditions far beyond the equilibrium melt line. Extensive analysis of experimental uncertainties including pressure gradients, off-Hugoniot states, shock unsteadiness, diagnostic sensitivity, and thermal diffuse scattering demonstrates that these effects cannot account for the magnitude of the observed discrepancy.

Another possibility, unconstrained by our measurements, that could broaden the apparent solid–liquid coexistence region on our rapid \textit{ns} timescales is the presence of kinetic barriers to melting. Under uniaxial shock compression, different portions of the shocked volume dwell at peak stress for different durations. Material near the ablation surface remains at high $P$-$T$ conditions for several nanoseconds, whereas material just behind the shock front experiences these thermodynamic conditions for only a fraction of a nanosecond. If the solid$\rightarrow$liquid transition in Ti is sluggish, we would expect the solid phase to persist to higher pressures than predicted by the equilibrium MLIP melt line. Related kinetic effects have been reported in Ge \cite{renganathan2023}. However, our data do not allow us to assess kinetic contributions directly.

Finally, we have not considered the possibility of more exotic melting behavior in Ti. Recent shock experiments on single-crystal aluminum \cite{renganathan2024,Renganathan2025} indicate that the melting pressure can depend on crystallographic orientation. An analogous orientation-dependent melting response in Ti would be consistent with our observations and could help explain the persistence of residual levels of solid $\beta$-Ti at pressures where complete melting is expected.

%We interpret the persistence of residual crystalline $\beta$-Ti as evidence for non-equilibrium melting behavior under rapid compression, potentially arising from orientation-dependent melting or kinetic hindrance in the solid--liquid transformation. Both mechanisms are consistent with the observed re-emergence of strong texture at the highest pressures and with recent reports of anisotropic melting in dynamically compressed crystalline solids. Our results therefore suggest that melting under shock loading cannot, in general, be described solely by equilibrium thermodynamics, even when advanced \textit{ab initio}--based models are employed.

These results have important implications for determining melting along the Hugoniot and for validating multiphase equations of state used in high–energy-density and planetary modeling. More broadly, this work highlights the intrinsic experimental challenges associated with identifying solid–liquid coexistence in laser-driven shock experiments. The determination of incipient melting is complicated by thermal diffuse scattering and potential laser-plasma heating, while the presence of epoxy bonding layers can influence the apparent completion of melting. In addition, drive unsteadiness impacts both the onset and completion of the inferred coexistence regime. Future experiments will be performed to investigate the effects of epoxy glue layer by replacing the Ti foil by a deposited layer. Potential time-dependent melting effects %which may give rise to the observed residual $\beta-$Ti grains at high pressure, 
will be investigated by probing the shock-compressed samples at late-time pressure-release states, to access the evolution of solid+liquid phase fractions. %This will address the question of how long lived the residual $\beta-$Ti grains are.
%will be investigating by , and on shock release to probe the spatial location of the residual $\beta-$Ti grains at high pressure. 

%microstructural sensitivity and phase-transition kinetics in interpreting dynamic compression experiments. Future studies employing controlled crystal orientations, variable shock dwell times, and time-resolved diffraction will be essential to disentangle anisotropic and kinetic effects and to establish a unified description of melting in materials subjected to extreme strain rates.

\section*{Acknowledgments}
\noindent
We would like to thank Carol Ann Davis for her help in preparing the Ti targets.~This work was performed under the auspices of the US Department of Energy by Lawrence Livermore National Laboratory under contract number DE-AC52-07NA27344 (LLNL-JRNL-XXXXXX). Use of the Linac Coherent Light Source (LCLS) at SLAC National Accelerator Laboratory is supported by the US Department of Energy, Office of Science, Office of Basic Energy Sciences under contract number DE-AC02-76SF00515. The MEC instrument is supported by the US Department of Energy, Office of Science, Office of Fusion Energy Sciences under the same contract.

\bibliographystyle{elsarticle-num}
\bibliography{references}
\onecolumngrid
%\title{Paper title}
\noindent
\begin{center}
    \textbf{\color{purple} \Large{SUPPLEMENTARY MATERIALS}}
\end{center}

\begin{center}
    
\bigskip
\Large{\textbf{High pressure melt dynamics in shock-compressed titanium}}

\bigskip
\noindent
\normalsize{Saransh Singh, Reetam Paul, Nikhil Rampal, Rhys J. Bunting, Sebastien Hamel, Nathan Palmer, \\Christopher P. McGuire, Samantha M. Clarke, Amy Coleman, Cara Vennari, Trevor M. Hutchinson, \\Kimberly A. Pereira, Bob Nagler, Dimitri Khaghani, Hae Ja Lee, Nicholas A. Czapla, Travis Volz, \\Ian K. OCampo, James McNaney, Thomas E. Lockard, Jon H. Eggert, Amy Lazicki, \\Christopher E. Wehrenberg, Andrew Krygier, Raymond F. Smith}

\end{center}

\renewcommand{\thefigure}{S\arabic{figure}}

\renewcommand{\thefigure}{S\arabic{figure}}
\renewcommand{\thetable}{S\arabic{table}}
\renewcommand\thesection{S\arabic{section}}
\renewcommand\thesubsection{S\thesection.\arabic{subsection}}

% \captionsetup[table]{name=Table,labelsep=period}
% \captionsetup[Table]{labelfont=bf}

\setcounter{table}{0}
\setcounter{figure}{0}
\setcounter{table}{0}
\setcounter{section}{0}
\normalsize

\bigskip
\bigskip
\noindent
%\Large{Supplementary Materials}

\normalsize

\bigskip
\noindent
\underline{\textbf{Supplementary Material Tables:}}

\bigskip
\noindent
\textbf{Table \ref{Table:Melt_Table}:} Calculated conditions of melt.

\noindent
\textbf{Table \ref{Table:titanium_summary_table}:} Summary of x-ray diffraction data.

\bigskip
\noindent
\underline{\textbf{Supplementary Material Figures:}}

\bigskip
\noindent
\textbf{Figure \ref{fig:histograms}:} Histogram of the dataset distributions.

\noindent
\textbf{Figure \ref{fig:parity}:} Parity plots comparing ML model predictions with DFT reference values.

\noindent
\textbf{Figure \ref{fig:FigMD}:} 
Final configuration from the MLMD melting simulation of BCC titanium

\noindent
\textbf{Figure \ref{fig:FigHugo}:} 
Comprehensive EOS plot showing multiple isochores along which MLMD simulations were performed.

\begin{figure}[!h]
\begin{center}
\includegraphics[width=1\columnwidth]{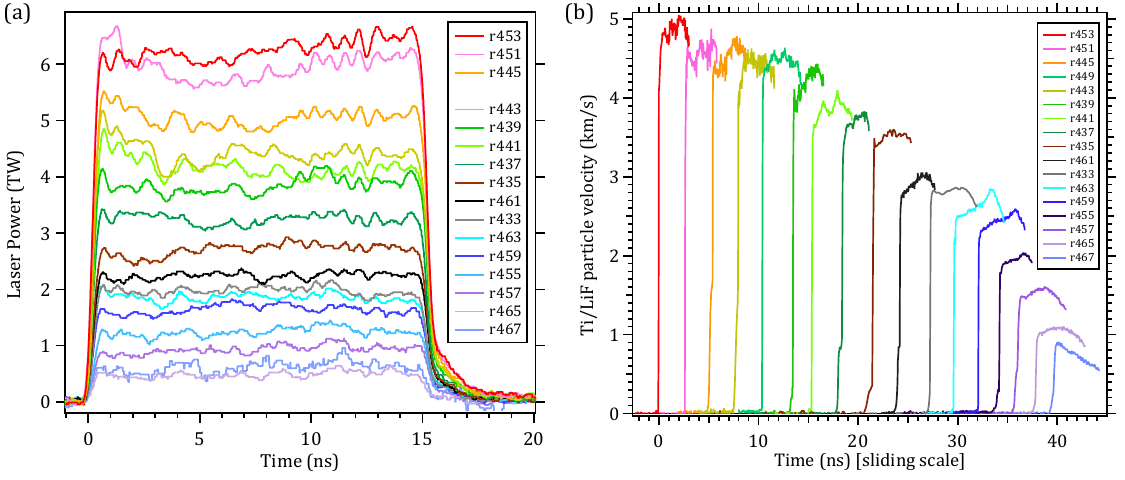}
%\captionsetup{width=0.5\linewidth}
\caption{\label{fig:Ti_Laser_Profiles} \textbf{(a)} Laser profiles for all shots within this study. \textbf{(b)} Velocity profiles for all shots within this study. 
}
\end{center}

\end{figure}

\noindent
\textbf{Figure \ref{fig:Ti_Laser_Profiles}:}
Laser profiles and measured Ti/LiF velocities for all shots within this study.

\begin{figure}[!h]
\begin{center}
\includegraphics[width=1\columnwidth]{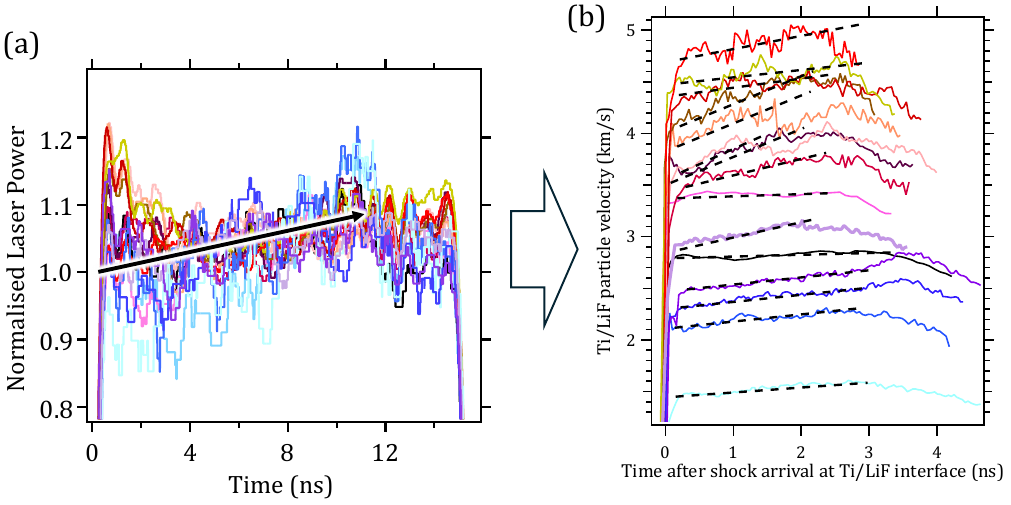}
%\captionsetup{width=0.5\linewidth}
\caption{\label{fig:Ti_Laser_Pulse_Shaping} \textbf{Sensitivity of laser pulse shaping to distribution of velocity states. (a)} Normalized laser power vs time for all fourteen $\sim$15-ns pulse shapes in this study. This shows the extent of pulse shape repeatability at MEC and, for this campaign, the gradual increase in laser power over the first $\sim$10.5-ns of the laser pulse shape.  \textbf{(b)} Corresponding Ti/LiF velocity profiles are characterized by a shock followed by ramp compression to peak pressure.
}
\end{center}

\end{figure}

\noindent
\textbf{Figure \ref{fig:Ti_Laser_Pulse_Shaping}:} 
Sensitivity of laser pulse shaping to distribution of velocity states.

\begin{figure}[!t]
\begin{center}
\includegraphics[width=1\columnwidth]{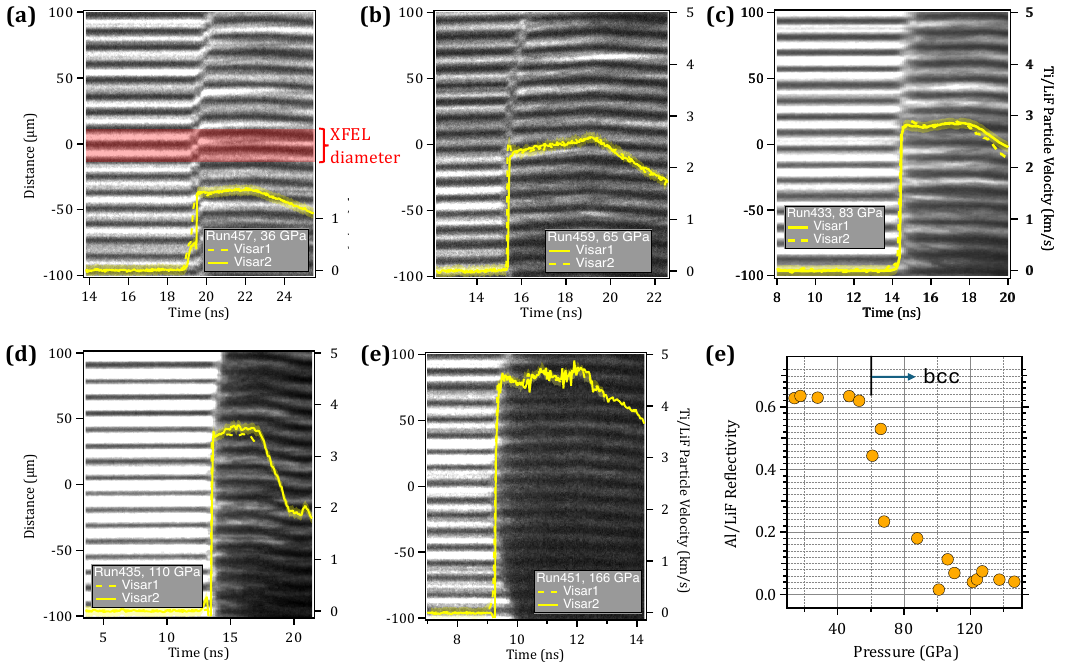}
%\captionsetup{width=0.5\linewidth}
\caption{\label{fig:Ti_VISAR_Reflectivity} \textbf{Summary of line-VISAR data.} Raw line-VISAR interferograms are shown for Ti shock pressures of \textbf{(a)} 36 GPa, \textbf{(b)} 65 GPa, \textbf{(c)} 83 GPa, \textbf{(d)} 110 GPa, and \textbf{(e)} 166 GPa. In all case the extracted Ti/LiF particle velocities are shown as the dashed and solid yellow curves for VISAR channel 1 and 2, respectively. There is a clear drop in Al/LiF reflectivity as a function of increasing pressure. \textbf{(e)} Reflectivity of the 0.2-µm thick Al layer coated onto the LiF window as a function of calculated Al pressure. There is a clear drop in the Al reflectivity at $P_{Al}$$\sim$50 GPa, which is equivalent to the pressure in Ti where bcc is first observed. We note that there is an expected drop in aluminum reflectivity with temperature \cite{mostovych1997}. In addition, the onset of a phase transformation can served to add spatial structure to the shock front which can reduce VISAR reflectivity \cite{smith2013b}.
}
\end{center}

\end{figure}

\noindent
\textbf{Figure
\ref{fig:Ti_VISAR_Reflectivity}:} Summary of line-VISAR data.

\begin{figure}[!t]
\begin{center}
\includegraphics[width=0.7\columnwidth]{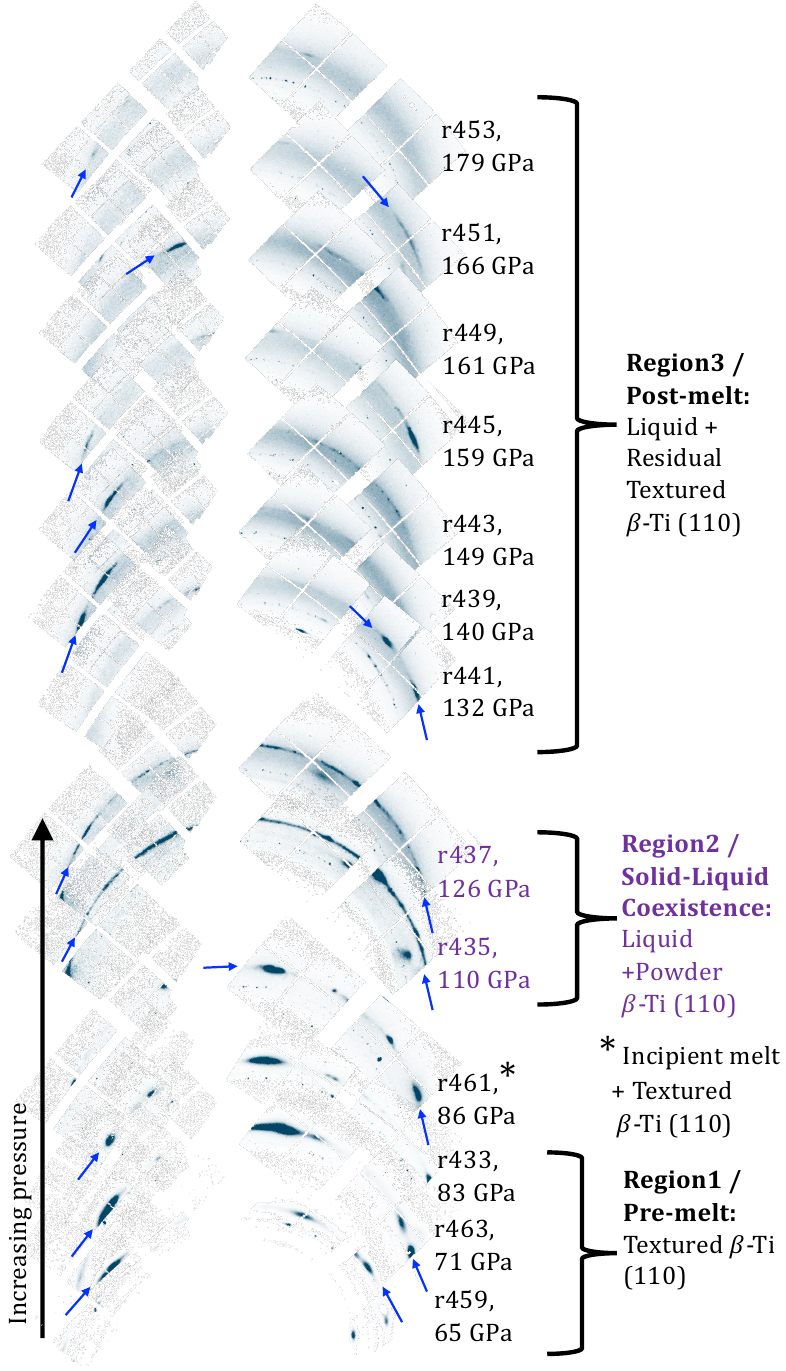}
%\captionsetup{width=0.5\linewidth}
\caption{\label{fig:XRD_Stereo_SM} \textbf{$\beta$-Ti X-ray diffraction patterns.} Select regions of x-ray diffraction pattern show the evolution of $\beta$-Ti (110) texture as a function of pressure. Same as for Fig. \ref{fig:Waterfall_texture}(b) except for all shots in this study.
}
\end{center}

\end{figure}

\noindent
\textbf{Figure
\ref{fig:XRD_Stereo_SM}:} $\beta$-Ti X-ray diffraction patterns.

\begin{figure}[!t]
\begin{center}
\includegraphics[width=0.7\columnwidth]{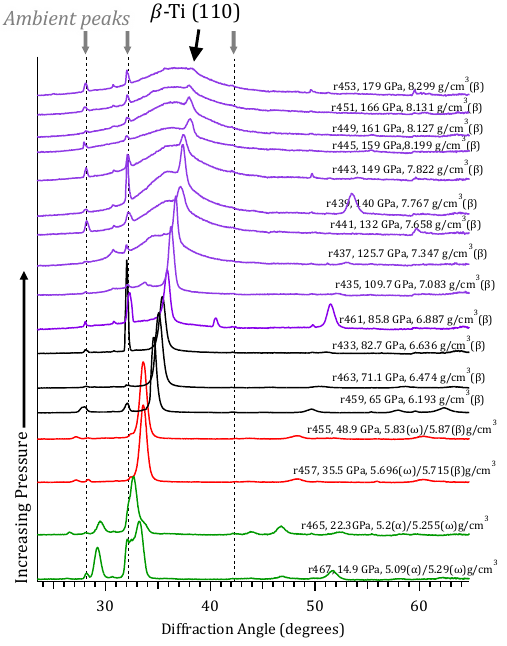}
%\captionsetup{width=0.5\linewidth}
\caption{\label{fig:Ti_Xray_profiles} \textbf{Ti x-ray diffraction profiles.} Azimuthally-averaged diffraction pattern as a function of increasing shock pressure.
Profiles are color-codes to reflect the structural identification: mixed $\alpha$+$\omega$ (green), mixed $\omega$+$\beta$ (red), $\beta$ (black), mixed $\beta$+liquid
(blue). Labels of experimental run number, and determined pressure from VISAR, and density per phase from XRD is labeled
on each trace. The position of the uncompressed peaks (originating from regions of the sample ahead of the shock front) are
highlighted by the vertical dashed lines. 
}
\end{center}

\end{figure}

\noindent
\textbf{Figure
\ref{fig:Ti_Xray_profiles}:} Ti x-ray diffraction data.

\begin{figure}[!t]
\begin{center}
\includegraphics[width=0.7\columnwidth]{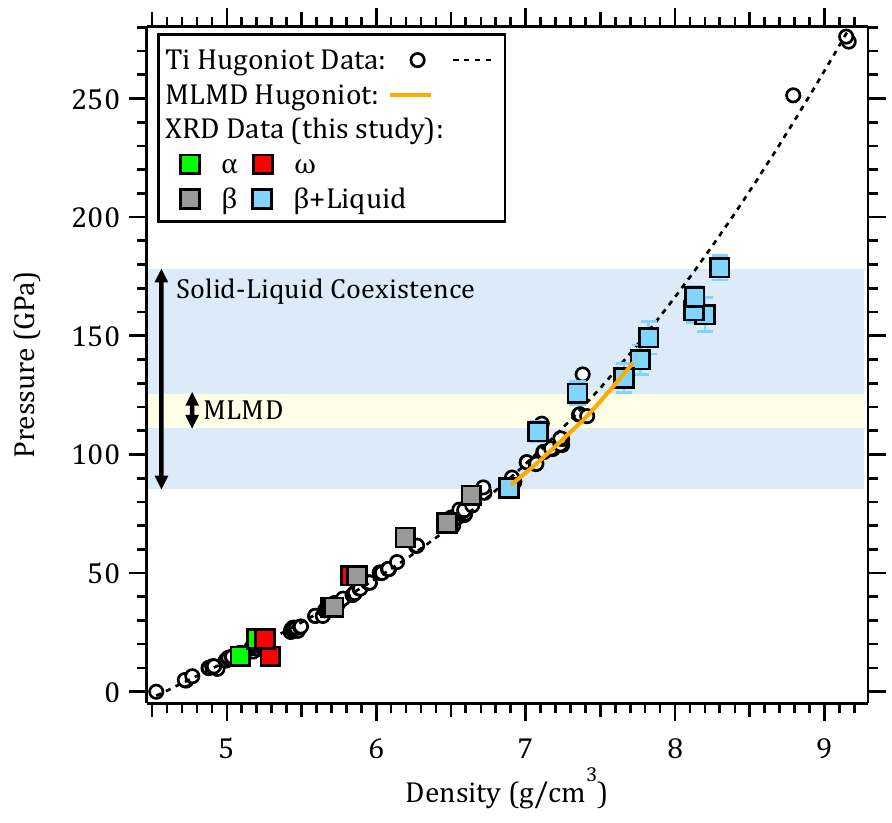}
%\captionsetup{width=0.5\linewidth}
\caption{\label{fig:Ti_Huogniot} \textbf{Ti Hugoniot Data.} Previously published titanium Hugoniot data are shown as white circles \cite{walsh1957,mcqueen1960,krupnikov1963, mcqueen1970,Marsh1980, al1981,trunin1999}. A fit to this data is shown as the dashed line. Data from out x-ray diffraction study are represented by the colored squares, color coded by the observed structural phase. Based on the phase fraction determination in the coexistence (Fig. \ref{fig:liquid_phase_fraction}), and the assumption that the phase fraction varies linearly in the coexistence region, we determine the solid liquid coexistence between $\sim$86-179 GPa (blue shaded region). The yellow shaded region shows the coexistence pressure determined by the MLMD simulations.
}
\end{center}
\end{figure}

\noindent
\textbf{Figure \ref{fig:Ti_Huogniot}:} Ti Hugoniot data.

\noindent
\textbf{Figure \ref{fig:liquid_phase_fraction}:} Determination of liquid phase fraction.

\noindent
\textbf{Figure \ref{fig:Ti_liquid_sensitivity}:} Minimum detectable volume of liquid Ti

\noindent
\textbf{Figure \ref{fig:view_factor}:} View factor between the laser spot and the XFEL probe.

\clearpage

\section{Obtaining the theoretical equation of state (EOS) for body-centered Titanium\label{sec:simulations_SM}} 
\squeezeup
\noindent
The EOS data was generated from theory using a multistage approach, as is the norm in modern materials science. Employing state-of-the-art machine learning (ML) approaches, we used a density functional theory (DFT)-generated dataset to train a machine learned interatomic potential (MLIP). This MLIP was thereafter used to perform molecular dynamics (MD) simulations in the \textit{P}-\textit{T} domain relevant to the experimental conditions. We discuss this step-by-step approach in detail below: 

\subsection{Generation of training data from DFT}
\begin{figure}[!t]
    \centering
    \includegraphics[width=0.8\linewidth]{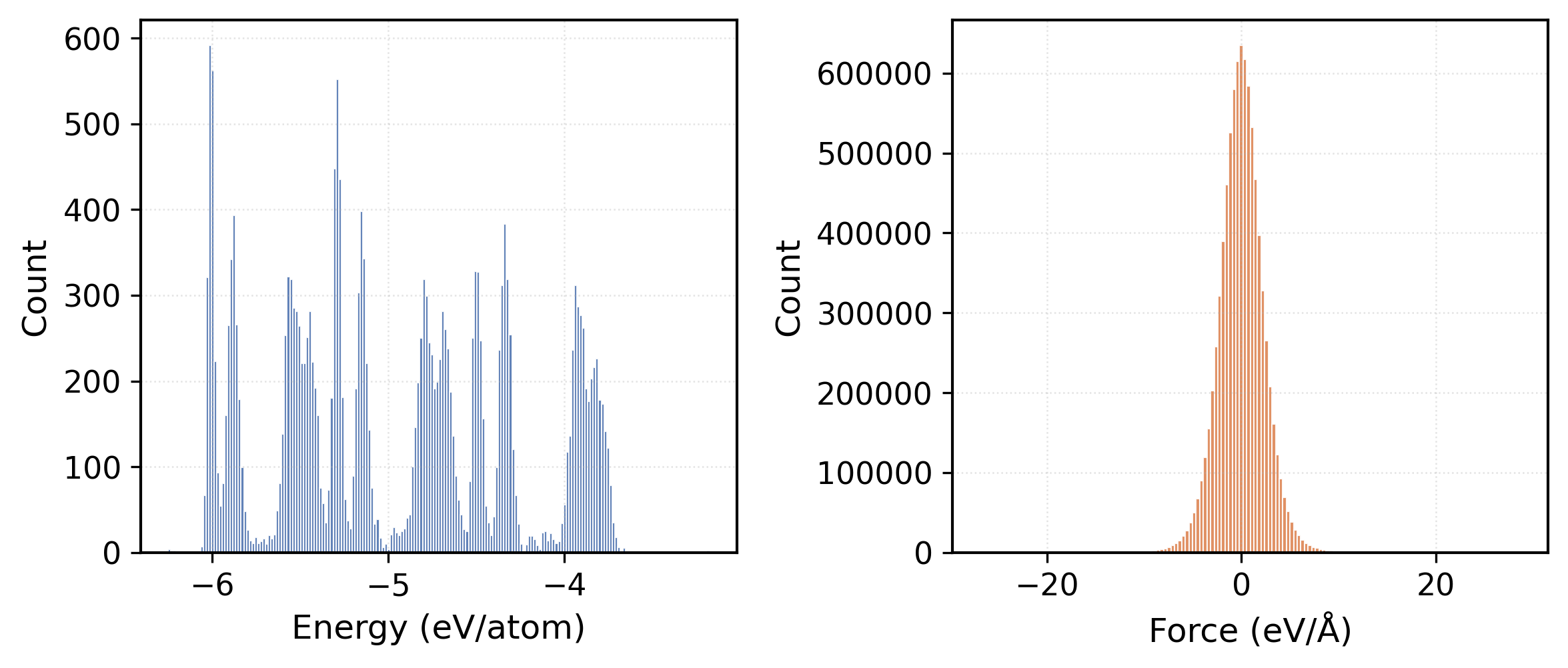}
    \caption{Histograms of the dataset distributions. Distributions of energy per atom (left) and force components (right) of the training dataset, spanning a broad range, reflecting the diversity of atomic environments sampled}
    \label{fig:histograms}
    \squeezeup
    \squeezeup
\end{figure}

\squeezeup
\noindent
The first principles data used for training was generated using \textit{ab initio} molecular dynamics (AIMD) simulations in the canonical (\textit{NVT}) ensemble, utilizing density functional theory (DFT) \cite{HohenbergKohn1964}, along multiple isochores. These DFT calculations used the Perdew-Burke-Ernzerhof (PBE) \cite{PerdewBurkeErnzerhof1996} exchange-correlation functional using the generalized gradient approximation (GGA) \cite{PerdewWang1992} as implemented in the Vienna \emph{ab initio} simulation package (VASP) \cite{Kresse1996}. For the DFT calculations, the cut-off energy of the plane wave basis set was \qty{500}{\eV}. The Brillouin zone was sampled using a Baldereschi (1/4,1/4,1/4) k-point mesh. For the AIMD simulations, Fermi-Dirac smearing was used to consider finite-temperature electronic effects, and temperature was set using the Nos\'e-Hoover thermostat\cite{nose1984unified, hoover1985canonical}. A $4 \times 4 \times 4$ body-centered cubic (BCC) Ti supercell of 128 atoms was used for these calculations to generate the dataset.
The initial dataset was generated from aforementioned AIMD simulations of \textbf{4} isochores simulated between densities ($\rho$) of 7.38 to 9.78 g/cm$^3$ (roughly mapping out 100 to 180 GPa), sweeping temperatures between 2000 to 4000~K for at least 9 ps at each ($\rho$, \textit{T}) point. The distribution of energies and forces in the final dataset are shown in Fig. \ref{fig:histograms}.

\subsection{Training of Machine-learned Interatomic Potential (MLIP)}
\begin{figure}[!t]
    \centering
    \includegraphics[width=0.8\linewidth]{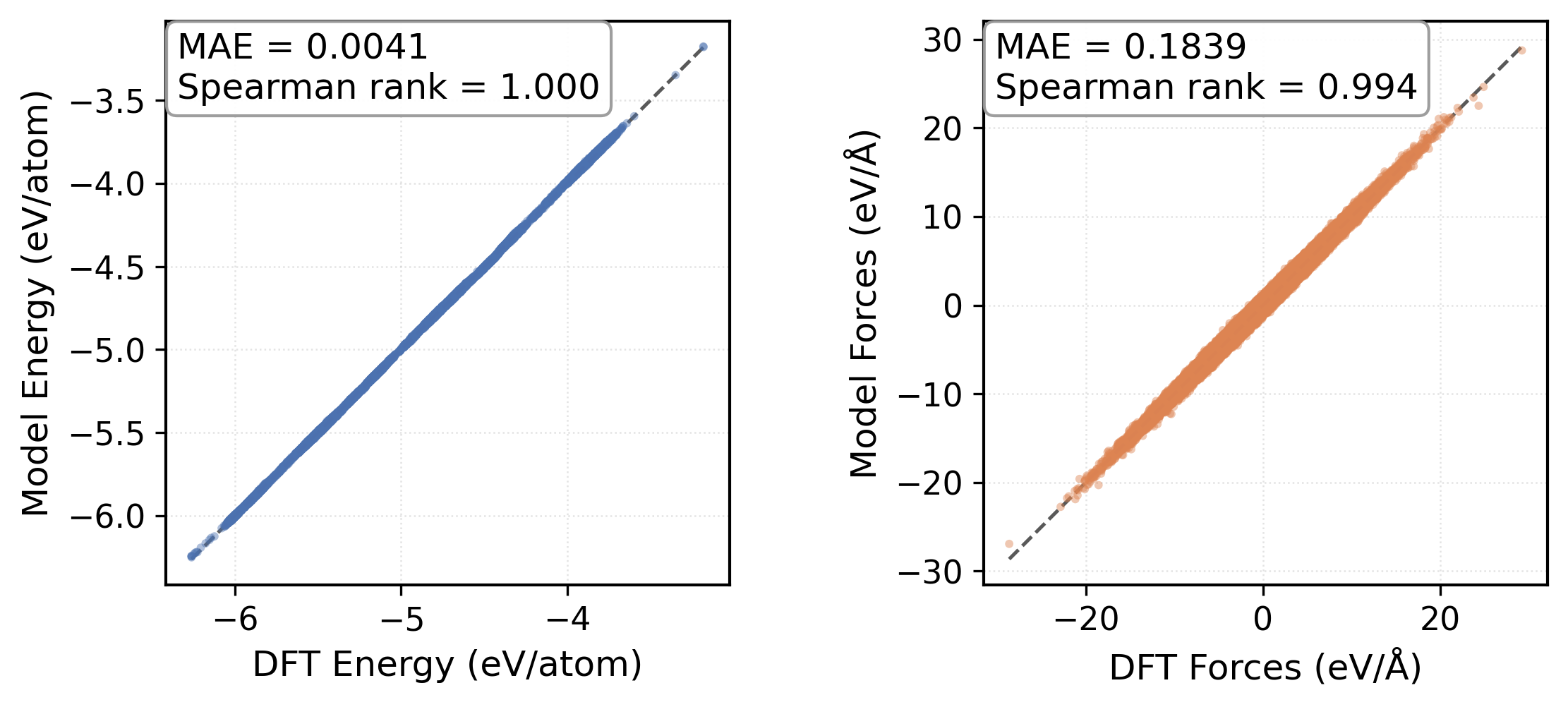}
    \caption{Parity plots comparing ML model predictions with DFT reference values. Parity of energy per atom (left) values and force components (right). The black dashed line represents perfect agreement between the ML model and DFT.}
    \label{fig:parity}
    \squeezeup
    \squeezeup
\end{figure}

\squeezeup
\noindent
The MLIPs were trained using the Allegro package\cite{musaelian2023learning,kozinsky2023scaling,tan2025high}. The atomic environments were described with a cutoff of $4.8$\AA~and $8$ Bessel functions with a polynomial envelope function of $p=5$. Here, $32$ features of even parity were considered, with a maximum rotation order truncation of $l_{max}=2$. Two layers were used in the network. The two-body latent MLIP had $2$ hidden layers of dimension $32$ per layer. The total network had $\sim14.2$k parameters. For training, $85$\% of the datasets were used for the training set, $10$\% for the validation set, and the final $5$\% were used for the test set. A batch size of $1$ was used. An initial learning rate of $0.001$ was used, with a cost function ratio of $10:1$ for force elements and total energies. The parity plots for energy and forces in the MLIP relative to the DFT values can be seen in Fig.~\ref{fig:parity}.

\subsection{Running machine-learned molecular dynamics (MLMD)}
\squeezeup
\noindent
We performed MLMD simulations using LAMMPS\cite{LAMMPS,plimpton1995fast} to investigate the melting behavior of BCC Ti. The initial configuration was constructed as a BCC lattice with a supercell size of $20 \times 20 \times 70$ unit cells along the $x$, $y$, and $z$ directions, respectively, with periodic boundary conditions being applied in all directions. Interatomic interactions, as described in (b), were modeled using the Allegro MLIP.

The simulation protocol began with equilibration at 1500~K using the Nos\'e-Hoover thermostat \cite{nose1984unified, hoover1985canonical} in the canonical (\textit{NVT}) ensemble for 50{,}000 steps. This was followed by further equilibration in the isothermal--isobaric (\textit{NPT}) ensemble at the same temperature and a target pressure ($P_{\mathrm{set}}$) for 100{,}000 steps. To explore pressure dependence, we performed a series of simulations with pressures ranging from 75~GPa to 180~GPa in 5~GPa increments, using isotropic pressure control and a barostat damping constant of 500~fs. The temperature damping constant was set to 200~fs for all \textit{NVT} and \textit{NPT} stages, and the timestep was 1~fs throughout all simulations.

\begin{figure}[t]
    \centering
    \includegraphics[width=0.25\linewidth]{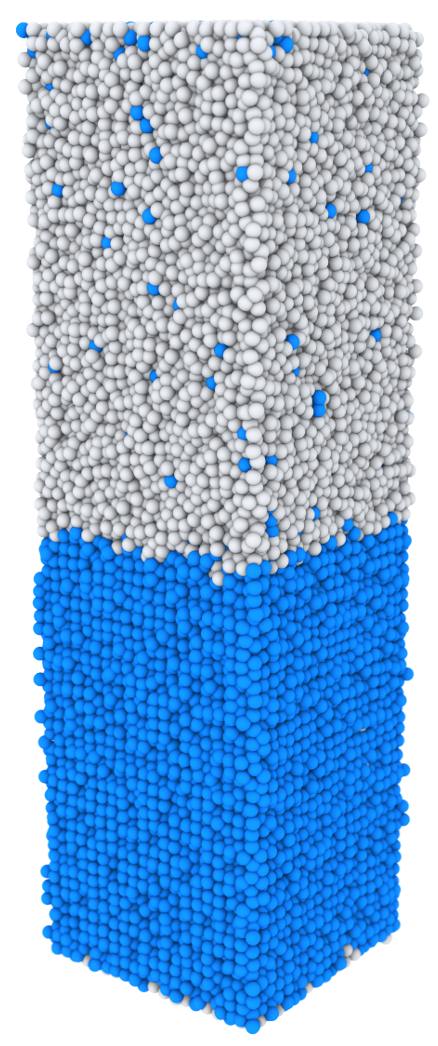}
    \caption{Final configuration from the MLMD melting simulation of BCC titanium on the melt-Hugoniot coexistence at 120~GPa, with atoms colored by the local Steinhardt $q_6$ order parameter. Atoms with $q_6 \geq 0.35$ are shown in blue, indicating crystalline environments characteristic of the solid phase, while atoms with $q_6 < 0.35$ are shown in gray, corresponding to the disordered liquid phase. The sharp transition between color regimes highlights the solid--liquid interface, with the bottom region predominantly solid and the top region liquid.}
    \label{fig:FigMD}
\end{figure}

To prepare a two-phase configuration for melting simulations, the equilibrated system was gradually heated from 1500~K to 100~K below the target melting temperature ($T_{\mathrm{melt}}$) over 100{,}000 steps under constant pressure. The simulation cell was then divided along the $z$-axis into two equal regions. The lower half, representing the solid phase, was frozen, while the upper half, representing the liquid phase, was heated from $T_{\mathrm{melt}} - 100$~K to $T_{\mathrm{melt}} + 2000$~K over 100{,}000 steps and subsequently cooled to $T_{\mathrm{melt}} + 100$~K over an additional 100{,}000 steps. The roles of the two regions were then reversed, and the same annealing protocol was applied to the previously frozen region. After both solid and liquid regions were equilibrated, the entire system was simulated at $T_{\mathrm{melt}}$ and $P_{\mathrm{set}}$ for 20{,}000 steps in the \textit{NPT} ensemble with anisotropic pressure control. This was followed by a long 1{,}000{,}000-step simulation in the \textit{NPH} ensemble, with pressure controlled only along the $z$-axis to enable observation of melting behavior under adiabatic conditions. Representative snapshot from the final trajectory are shown in Fig.~\ref{fig:FigMD}. 

In Fig.~\ref{fig:FigMD}, atoms are colored by their local Steinhardt $q_6$ order parameter, providing a continuous measure of local structural order and highlighting the sharp transition across the solid--liquid interface. With $q_6$ being normalized, as is our case, $ 0 \le q_6 \le 1 \,$. $q_6=1$ occurs only in the degenerate limit where all bond vectors are parallel. For isotropic (random) bond orientations, $q_6 \to 0$, for BCC with 8 nearest neighbors, $q_6 \approx 0.51069$. 

Compilation of all the pressure(\textit{P})-temperature(\textit{T})-density ($\rho$)-internal energy(\textit{U})-free energy (\textit{H}) data from the MLMD simulations map out a comprehensive equation of states (EOS) dataset that can be used to aid in analysis of broader trends in the experimental data.

\bigskip

(d) \textbf{Calculation of principal Hugoniot and isentropes from the equation of state data}:

Once the EOS data was generated from the MLMD simulations, the Hugoniot and isentropes are derived using the standard mathematical formulation as described below.

\begin{figure}[!t]
\begin{center}
    \includegraphics[width=1\linewidth]{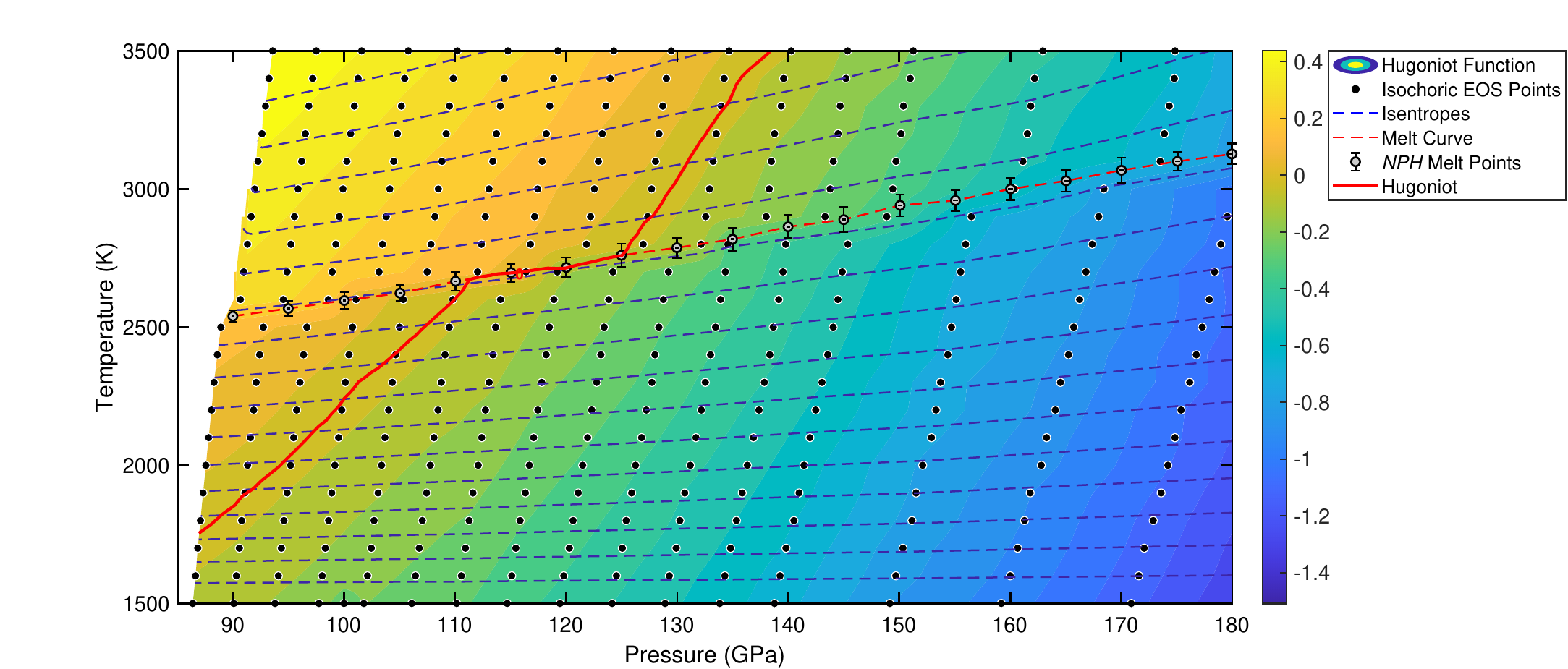}
    \caption{Comprehensive EOS plot showing multiple isochores along which MLMD simulations were performed. The Hugoniot function is shown as a contour plot, and the locus of zero values of this Hugoniot function is shown here as the \textit{principal} Hugoniot (red solid line). Multiple isentropes are plotted with different launch points on the first isochore. The \textit{NPH} melt points are shown with uncertainties, and a melt curve is also shown.}
    \label{fig:FigHugo}
\end{center}

\end{figure}

The Hugoniot function, defined using the Rankine-Hugoniot relation as
\begin{equation}
H(P, T)=E(P, T)-E_0-\frac{1}{2}\left[P+P_0\right]\left[V_0-V(P, T)\right],
\end{equation}
was calculated directly from the MLMD data. The locus of points corresponding to $H(P, T)=0$ is the principal Hugoniot and is shown in Fig.~\ref{fig:FigHugo} below as a solid red line.

For the isentropes, we start by defining the total differential of the specific
entropy for a single-component system: 
\[
  dS=\frac{\partial S}{\partial T}\,dT+\frac{\partial S}{\partial v}\,dv,
\]
where \(v\) is the specific volume. We invoke the following Maxwell relations
\begin{equation}
\begin{aligned}
& \frac{\partial S}{\partial T}=\frac{1}{T} \frac{\partial E}{\partial T } \, \, \text{and}  \, \,  \frac{\partial S}{\partial V}=\frac{\partial P}{\partial T}.
\end{aligned}
\end{equation}
Using the first law substitution, \(dE = T\,dS - P\,dv\) and rearranging for an isentrope (\(dS=0\)) gives
\[
  \frac{1}{T}\,\frac{\partial E}{\partial T}\,dT
  \;=\;
  -\frac{\partial P}{\partial T}\,dv .
\]
Dividing the previous expression by
\(\frac{\partial E}{\partial T}\) yields  
\[
  \frac{1}{T}\,dT
  \;=\;
  -\frac{\frac{\partial P}{\partial T}}
         {\frac{\partial E}{\partial T}}\;dv
  \;=\;
  -\!\left[\frac{\partial P}{\partial E}\right]_V dv
\]
allows the calculation of the locus of points that define an isentrope while hopping from one isochore to another.

Fig.~\ref{fig:FigHugo} shown below is a composite plot of the pressure-temperature EOS with principal Hugoniot, multiple isentropes starting from different launch points on the first isochore, and melt curve (fine-tuned using \textit{NPH} simulations) superimposed. As can be seen, the slopes of the melt curve and isentropes (near melt) are approximately equal for a wide expanse of \textit{P}-\textit{T} conditions. Hugoniot-melt coexistence spans a pressure range of $\sim$111 to $\sim$124 GPa. 

\clearpage

\begin{table}[!t]
\begin{center}
    \includegraphics[width=0.75\linewidth]{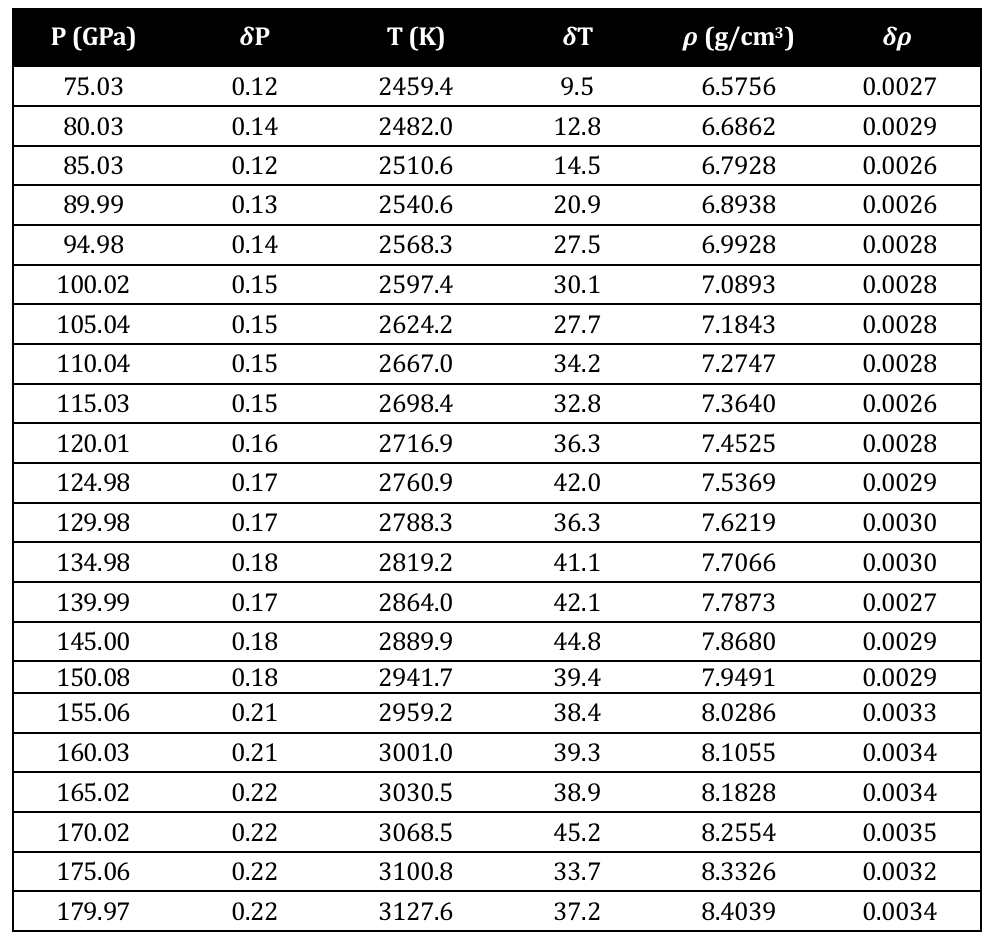}
    \caption{Calculated conditions of melt}
    \label{Table:Melt_Table}
\end{center}

\end{table}

\clearpage

\section{Pressure Determination} 
\noindent
To analyze the VISAR data to provide a constraint on sample pressure we follow the approaches put forth in Refs. \cite{Volz2026, McGuire2025}. Non-ideal drive conditions will result in a range of pressure states within the sample. The extent to which pressure varies in time and with distance across the compressed sample volume is captured by variations within the VISAR record of Ti/LiF particle velocity with time, $u_P$(t). Following the procedure outlined in Ref. \cite{Volz2026} we use the initial $u_P$ after shock arrival at the Ti/LiF interface to get an initial estimate of sample pressure based on standard impedance-matching techniques \cite{pascarelli2023} using a knowledge of Hugoniot relations for Ti \cite{walsh1957,mcqueen1960,krupnikov1963, mcqueen1970,Marsh1980, al1981,trunin1999} and LiF \cite{hawreliak2023, rigg2014}.~To calculate how much (in time) of the post-shot $u_P$(t) data is relevant for a final estimation of pressures we consider the expected pressure-dependent shock speed in the Ti sample \cite{walsh1957,mcqueen1960,krupnikov1963, mcqueen1970,Marsh1980, al1981,trunin1999} and the x-ray probe time relative to the shock entry time into the 32-µm-thick Ti foil. Over this period a weighted average of the measured $u_P$(t) trace from two independent VISAR measurements, which considers the uncertainty in each measurement, provides an average value of $u_P$ and $\sigma_{u_P}$.

To determine the pressure and pressure uncertainty in the Ti we follow the approach outlined in Ref.~\cite{McGuire2025}. Measurement errors in $u_p$ were propagated using a Monte Carlo method through reflected-shock impedance matching \cite{pascarelli2023} along with the associated uncertainties in the EOS fitting parameters of LiF \cite{rigg2014}, and Ti \cite{walsh1957,mcqueen1960,krupnikov1963, mcqueen1970,Marsh1980, al1981,trunin1999}. The estimated pressures and uncertainty in pressure is shown in Table \ref{Table:titanium_summary_table}.

\begin{table}[!h]
\begin{center}
    \includegraphics[width=0.95\linewidth]{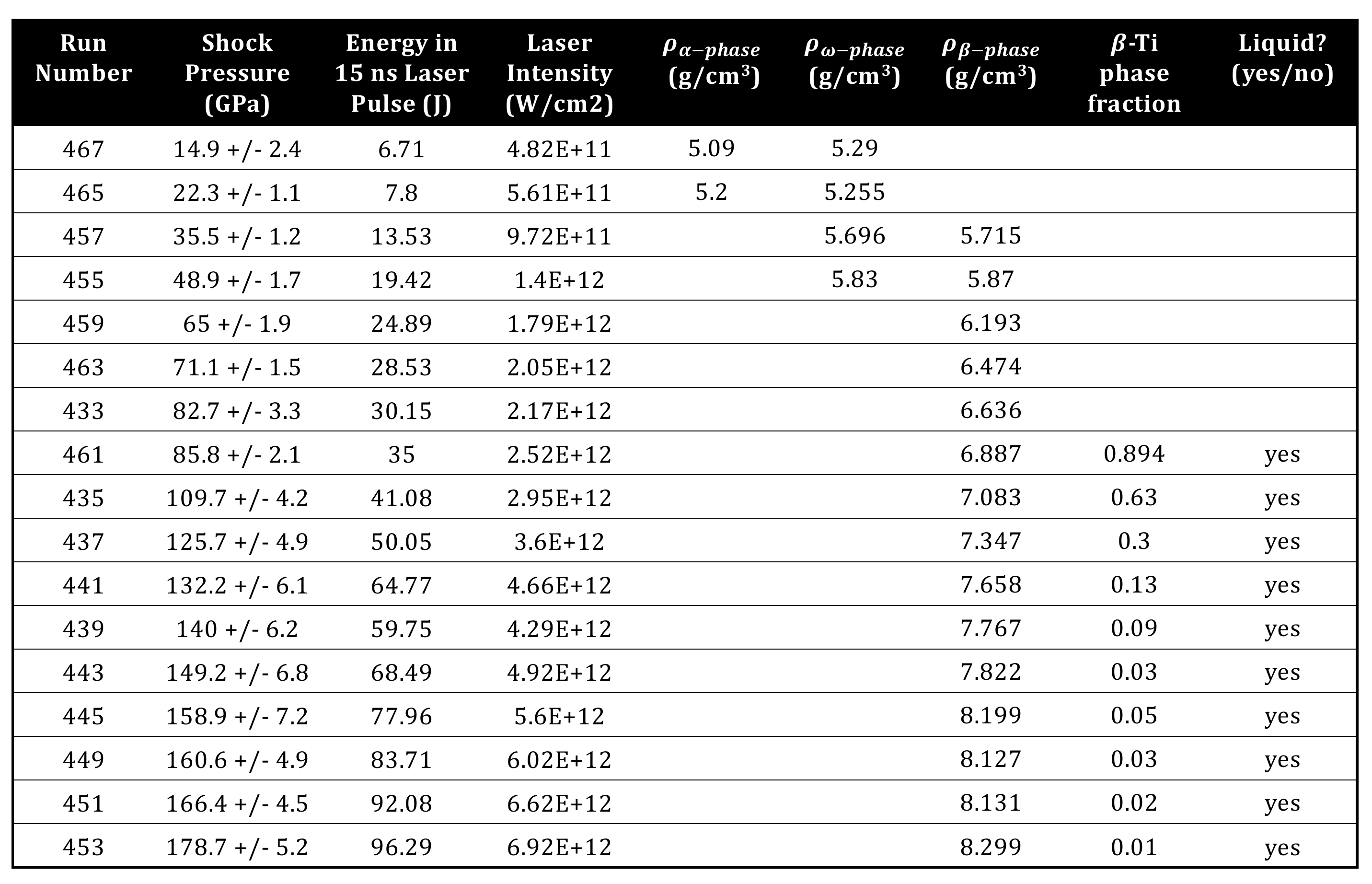}
    \caption{Summary of x-ray diffraction data.}
    \label{Table:titanium_summary_table}
\end{center}

\end{table}

\clearpage

\clearpage

\clearpage

\clearpage

\clearpage

\clearpage 

\section{Liquid phase fraction as function of pressure}
\noindent
Here, we present a model for extracting the phase fraction of the liquid Ti in the pressure range with $\beta-$Ti-liquid coexistence. We utilize the pseudo-amorphous approximation within the quantitative phase analysis method \cite{Lebail1995,Lutterotti1998}. Since the liquid phase emerges from the bcc $\beta-$phase, we use a combination of two bcc phases to perform a Rietveld refinement. The first bcc phase models the diffraction from the solid phase and the second bcc phase models diffraction from the liquid phase. The diffuse nature of the liquid diffraction signal was modeled by extreme peak broadening of the pseudo-amorphous bcc phase. We also included a thermal diffuse scattering model presented in references \cite{warren1990,heighway2019} for the incipient melt shots. The TDS model accounts for the small amount of thermal diffuse scattering signal from the $\beta-$Ti phase. The background was modeled using a linear combination of low order (up to 2$^{\textrm{nd}}$ degree) Chebyshev polynomials. There are two main sources of errors in the methodology applied here
\begin{enumerate}[wide, labelwidth=!, labelindent=0pt]
    \item Preferred orientation or crystallographic texture: In the presence of texture in the solid phase, the intensity of the solid signal is modified from that of the pure powder sample. This can lead to an incorrect estimate of the phase fraction in the diffracting volume. The estimates can be higher or lower depending on the nature of the texture. We account for this effect by including a spherical harmonic texture model \cite{VonDreele1997}.
    \item Insufficient grain sampling statistics in the scattering signal: When an insufficient number of grains scatter in the solid phase, the measured intensity is not representative of the actual solid volume. This can lead to an underestimate of the solid phase fraction. This effect is not taken into account in our estimates.
\end{enumerate}

\begin{figure}[!h]
\begin{center}
\includegraphics[width=0.8\columnwidth]{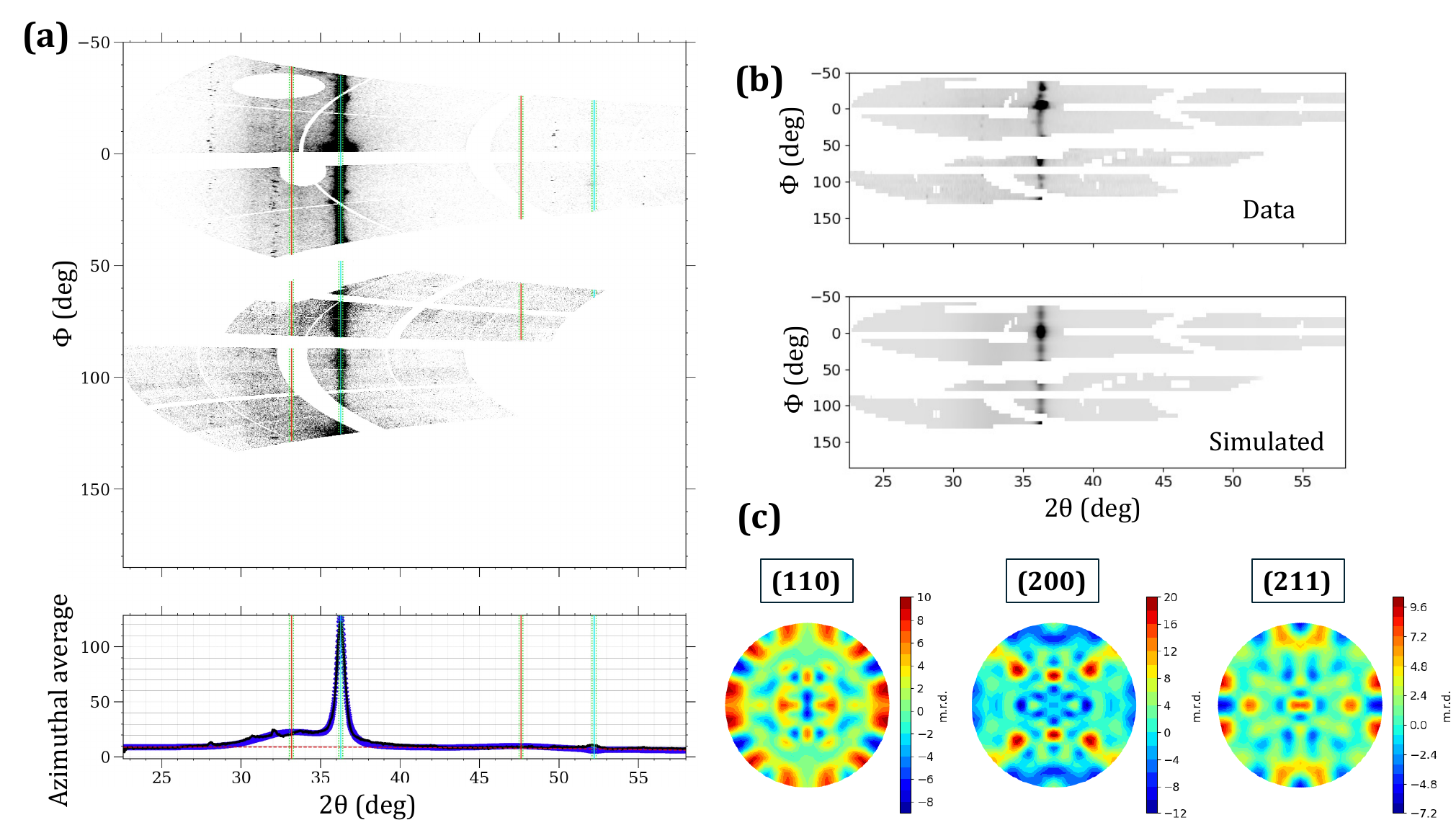}

\caption{\label{fig:liquid_phase_fraction} \textbf{Determination of liquid phase fraction in Ti with preferred orientation.} The phase fraction in the solid-liquid coexistence is determined by the traditional Rietveld method. Since the bcc $\beta-$phase melts to form the liquid, we model diffraction from the liquid phase by the bcc $\beta-$phase with extreme peak broadening \cite{Lebail1995,Lutterotti1998} and no preferred orientation. The texture in the solid $\beta-$phase was modeled using a generalized spherical harmonic basis functions \cite{VonDreele1997}. \textbf{(a)} Rietveld fit to the 1D azimuthally averaged diffraction signal, including the liquid and preferred orientation model for run 435.  \textbf{(b)} Experimentally recorded and simulated 2D diffraction patterns using the same liquid and texture models, and \textbf{(c)} simulated pole figures for the first three diffracting planes of the bcc $\beta-$phase. The Rietveld fit was performed using the HEXRD package \cite{HEXRD2025}.
}
\end{center}
\end{figure}

Our high-pressure data has both preferred orientations (intermediate pressure) and, in some cases, insufficient grain sampling statistics (high-pressures). In addition, in most of our high-pressure shots, the diffraction signal from the high-pressure solid $\beta-$phase has signal only from the $(110)$ reflection. The presence of just one diffraction line increases the uncertainty of correct texture determination. While the preferred orientation can be accounted for using this method, we are unable to model the uncertainties due to insufficient grain sampling statistics at higher pressures. All these compounding effects are reflected in the significantly higher uncertainties for all phase-fraction estimates ($\sim 10\%$) in our higher-pressure experiments compared to the traditional uncertainty estimates of $\sim 1\%$ for this method. We show the results of the fit, including preferred orientation for one of the shots (run 435) in Fig.~\ref{fig:liquid_phase_fraction}.

\clearpage

\section{Minimum detectable liquid volume\label{sec:detectability}}
\begin{figure*}[!t]
\begin{center}
\includegraphics[width=0.8\textwidth]{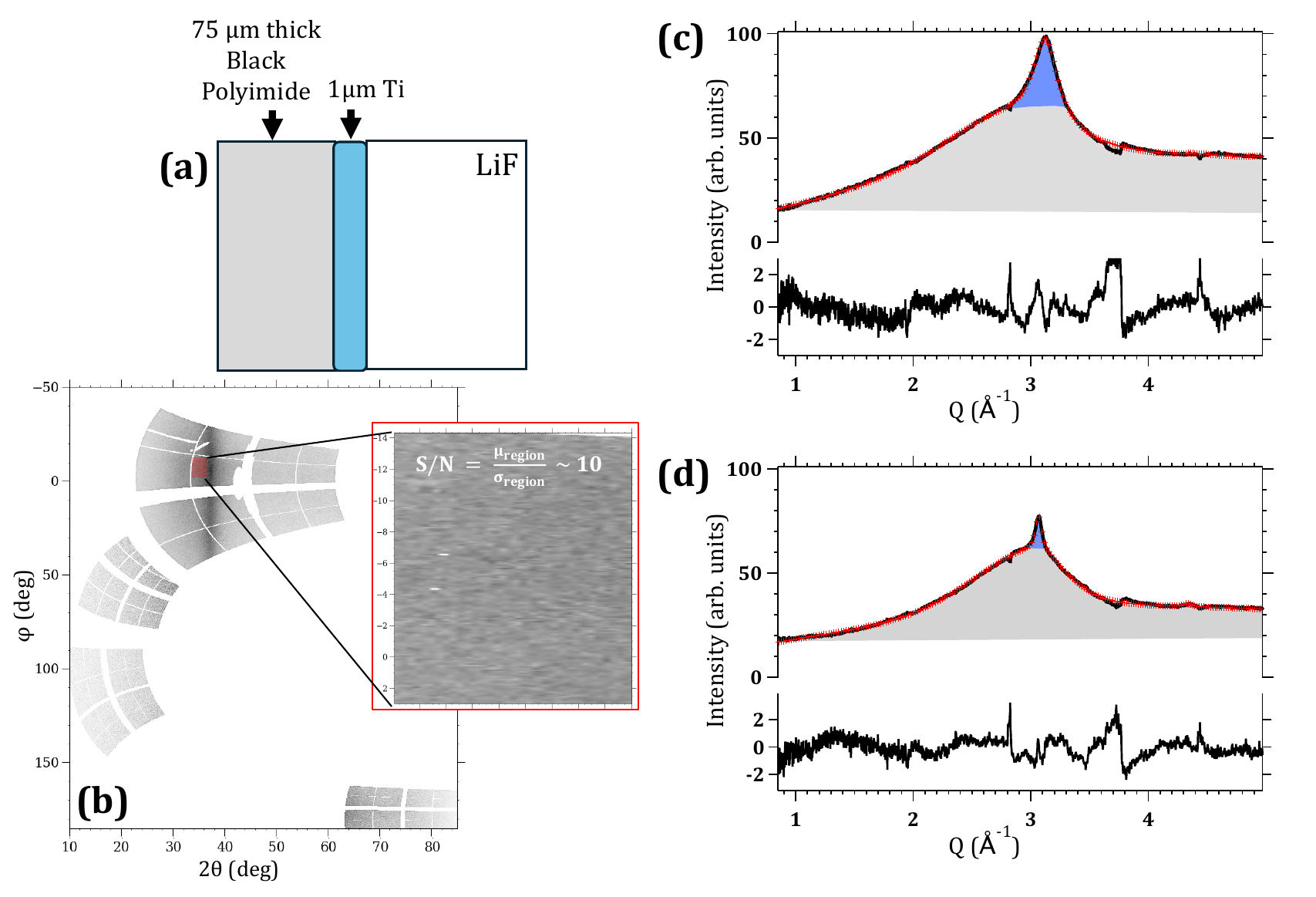}
% \captionsetup{width=0.5\linewidth}
\caption{\label{fig:Ti_liquid_sensitivity} \textbf{Minimum detectable volume of liquid in Ti.} We performed two diffraction experiment with a 1-$\mu$m-thick Ti sample. The sample was shocked to $\sim130$ GPa and released. Subsequently, x-ray diffraction data was recorded on the epix detector. The diffraction signal was consistent with a combination of bcc  and liquid Ti for all of the experiments. Liquid phase fraction was determined using two different methods: (i) degree of crystallinity (DOC) which measured the integrated areas under the peaks of solid and liquid phase and (ii) traditional Rietveld method with extreme peak broadening to model the liquid diffraction. Both of these methods agreed to within 1\%. The solid signal is shaded by blue and the liquid signal is shaded in gray \textbf{(a)} Target design consisting of a layer of polyimide/Ti/LiF. The 1-$\mu$m-thick Ti layer was deposited directly on the LiF substrate. \textbf{(b)} We estimate the signal-to-noise (S/N) ratio in these experiments by taking the ratio of the mean to the standard deviation of the diffraction signal from a region with the liquid diffraction peak. The signal-to-noise for the shot with the lower liquid fraction was determined to be $\sim10$ using this method. We conservatively assume that a liquid signal can be differentiated from noise till S/N $\geq 2$. \textbf{(c)} Results of the best fit with the traditional Rietveld method. The difference curve is shown at the bottom. We determined the liquid phase fraction to be $85\%$ for this sample. \textbf{(d)} results of the best fit with the DOC method. The difference curve is shown at the bottom. We determined the liquid phase fraction to be $97\%$ for this sample. Based on the data recorded in these experiments, we conclude that the minimum detectable volume of liquid Ti in thicker samples with 10 keV X-rays using the epix detector is approximately $0.85 \mu$m. Readers are referred to the text for detailed calculation supporting this estimate.
}
\end{center}

\end{figure*}

\noindent
We performed in situ X-ray diffraction on $1\ \mu$m thick shock compressed and released Ti. The target design, shown in Fig.~\ref{fig:Ti_liquid_sensitivity}(a), consists of a 75-$\mu$m-thick polyimide ablator, a 1-$\mu$m-thick Ti layer, and a 150-$\mu$m-thick LiF substrate. The Ti layer was deposited directly onto the LiF. We conducted two experiments of this type to constrain the minimum detectable volume of liquid Ti using 10 keV X-rays recorded on the ePix detectors. The Ti layer was shock compressed to peak pressures of approximately $85$ and $133$ GPa, respectively. The corresponding azimuthally averaged lineouts are shown in Figs.~\ref{fig:Ti_liquid_sensitivity}(a) and \ref{fig:Ti_liquid_sensitivity}(b). The diffraction signal in both shots is consistent with a mixture of solid bcc $\beta$ phase and liquid Ti.

We estimated the liquid phase fraction in the compressed volume using two methods: (i) a degree of crystallinity (DOC) analysis, which evaluates the integrated areas under the solid and liquid contributions, and (ii) a traditional Rietveld refinement with extreme peak broadening to represent the liquid diffraction. In the DOC method, the liquid diffraction signal is modeled using an asymmetric pseudo-Voigt peak shape, whereas in the Rietveld analysis it is represented by a bcc $\beta$ phase with very large peak broadening. The two approaches agree to within 1\%. From these measurements, the liquid thickness is estimated to be $\sim 0.85\ \mu$m. For the measurements described here, the approximate signal-to-noise ratio (S/N) was obtained by taking the ratio of the mean to the standard deviation of the diffraction intensity in regions near the liquid diffraction peak. We estimate $\text{S/N} \sim 10$ (equivalent to $\sim 10$ dB) for these experiments (see Fig.~\ref{fig:Ti_liquid_sensitivity}(b)). 

% Assuming that the liquid diffraction signal remains detectable down to a S/N ratio of $\sim 2$, and that the S/N decreases linearly with liquid thickness, we estimate the minimum detectable liquid thickness to be $\sim 0.85/5 \sim 0.19\ \mu$m.

In experiments with thicker Ti samples, the minimum detectable liquid fraction is expected to be higher, due to self-absorption of the liquid diffraction signal in Ti. Therefore, the $0.19\ \mu$m minimum liquid thickness detectable in the present configuration should be regarded as a best-case scenario, that is, a lower limit. We extend this estimate to thicker Ti targets by calculating the attenuation of the liquid signal as a function of Ti thickness, while keeping the noise level fixed.

 Additional Ti thickness reduces the signal but not the noise, leading to a lower S/N. Assuming the liquid Ti is uniformly distributed throughout the Ti sample thickness, the decay in the signal, $D$, for a sample of thickness $t$ at a scattering angle $2\theta$ can be approximated by
\begin{equation*}
    D(t) = \frac{\textrm{e}^{-\mu t} - \textrm{e}^{-\mu t\sec 2\theta}}{\mu t(\sec 2\theta - 1)}.
\end{equation*}
Here, $\mu$ is the attenuation coefficient of the sample. This expression accounts for the attenuation of the X-rays (i) as they travel normal to the sample for some depth before encountering the scattering volume and (ii) travel at an angle ($2\theta$) after scattering for the rest of the sample thickness. At 10 keV, Ti has an attenuation coefficient of $\sim 0.051\ \mu\textrm{m}^{-1}$. Using this value, together with a Ti thickness of 32 $\mu$m and a scattering angle of $\sim 32^{\circ}$ (corresponding to the peak of the liquid diffraction), we obtain a decay factor $D \sim 0.17$. Thus, self-absorption in our experimental geometry is expected to reduce the S/N to $\sim 1.7$, which is just below the detectability threshold of $S/N\sim2$. Therefore, we treat $0.85\mu$m as the minimum detectable liquid thickness in our experiments. We note that the LiF window used in our experiments also absorb some of the scattered photons. However, we ignore those because of the small absorption cross-section of LiF at 10 keV.

% Again, assuming that the signal level scales linearly with the thickness of the liquid layer, then the minimum detectable liquid thickness corresponding to a S/N of $\sim 2$ is given by
% \[
%     \frac{2 \times 0.85\ \mu\textrm{m}}{4.4} \sim 0.4\ \mu\textrm{m}.
% \]

% The assumption that the liquid is distributed uniformly over the shocked volume leads to an overestimate of the minimum detectable liquid layer thickness. In a planar shock geometry, sample layers closer to the shock front have been compressed for a shorter time. Therefore, if the solid $\rightarrow$ liquid transition is kinetically hindered, a larger fraction of the liquid will localize closer to the shock front. In this scenario, the liquid diffraction signal experiences less self-attenuation within the sample, which improves the detectable liquid volume.

\section{View factor analysis\label{sec:view_factor}}
\begin{figure*}[!t]
\begin{center}
\includegraphics[width=0.7\textwidth]{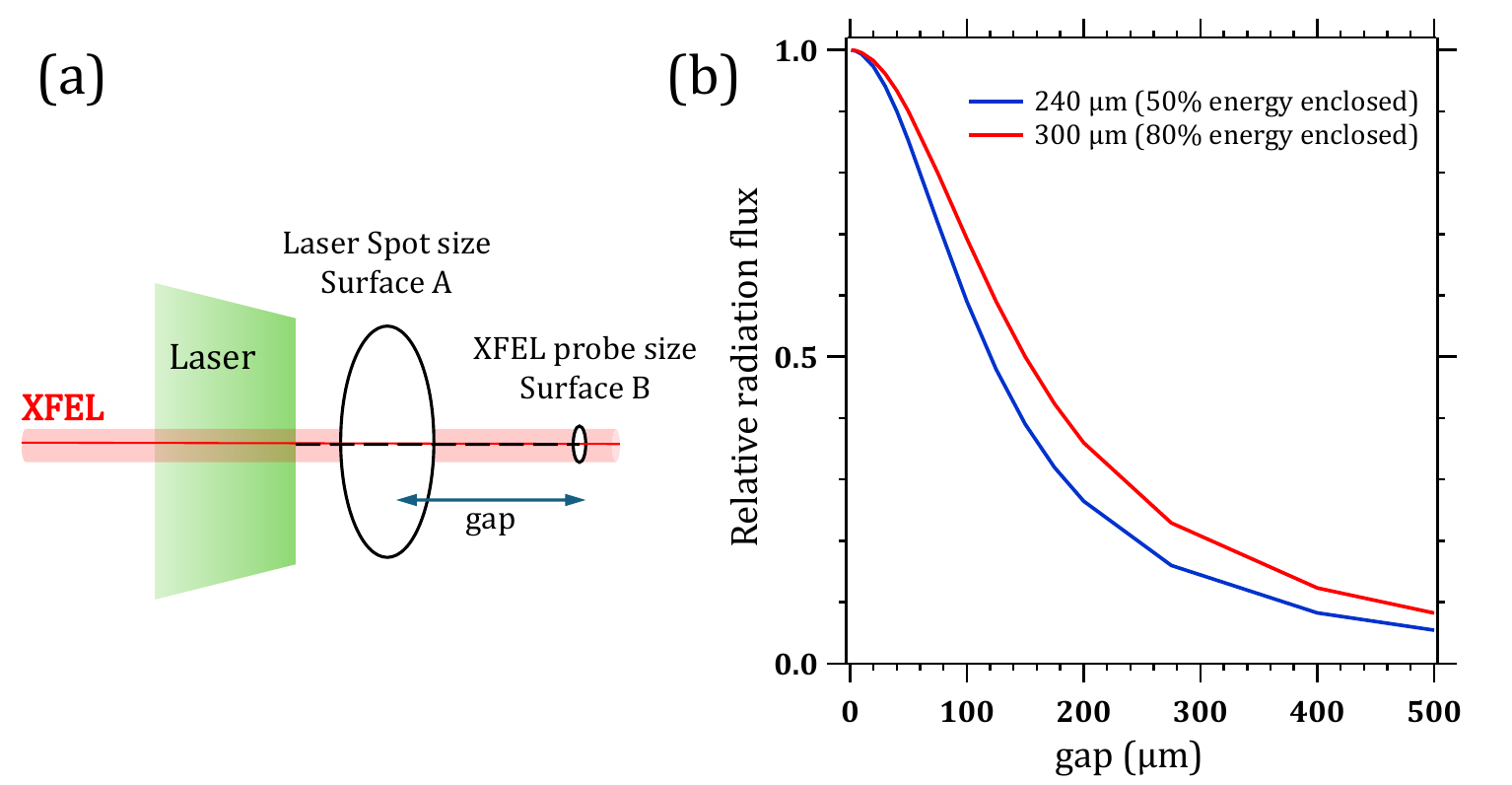}
% \captionsetup{width=0.5\linewidth}
\caption{\label{fig:view_factor} \textbf{View factor between laser spot and XFEL probe.} (a) Radiation leaving surface $A$ (laser spot) is intercepted by surface $B$ (XFEL probe). Both surfaces are approximated as discs. It is also assumed that the intensity of the radiation leaving surface $A$ (laser spot) is uniform over the surface. This is a good approximation since the laser spot is a super-gaussian with a high index. (b) Relative view factor, normalized to the zero-gap value, as a function of gap thickness for two laser spot sizes. For our heating experiments, the gap was $200~\mu\mathrm{m}$, corresponding to a relative view factor of $0.26$--$0.36$.
}
\end{center}

\end{figure*}

\noindent
Our laser plasma heating experiments described in detail in the main text (Section~\ref{sec:plasma_heating} and Fig.~\ref{fig:Xray_heating}) measures the heating in a $2.5~\mu$m thick Ti sample which using X-ray diffraction. There is a $200~\mu$m standoff between the ablator and the sample to isolate the laser plasma induced heating from the compression induced heating. To extend the validity of these measurement to our primary Ti data where there is effectively no gap between the ablator and the sample, we scale the heating by the view factor \cite{Siegel1992}, $F_{A\rightarrow B}$. The factor computes the radiation captured by surface $B$ leaving surface $A$. In our case, both the Laser spot (surface $A$) and XFEL probe (surface $B$) can be approximated as discs (see Fig.~\ref{fig:view_factor}(a). The view factor is given by the expression
\begin{align*}
    F_{A\rightarrow B} &= \frac{x-y}{2}, \nonumber \\
    x &= 1 + H^2/R^{2}_A+R_{B}^2/R_{A}^2, \nonumber \\
     y &= \sqrt{x^2-4R_{B}^2/R_{A}^2}.
\end{align*}
Here, $R_A, R_B$ and $H$ denote the laser spot size, the XFEL probe size and the gap thickness respectively. In our experiments using the $300~\mu$m continuous phase plates, $50\%$ of the laser energy is enclosed in a radius of $\sim240~\mu$m and $80\%$ is enclosed in a radius of $\sim300~\mu$m. The XFEL probe size in our experiments were $\sim20~\mu$m in size. Figure~\ref{fig:view_factor}(b) presents the view factor as a function of the gap thickness for the two laser spot sizes. The values are normalized with respect to the zero gap thickness.

For our heating experiments, the gap thickness was $200~\mu$m. The relative view factor for this thickness is $0.26$ and $0.36$ for the two laser spot sizes. Therefore, the heating in our primary experiment will be approximately $2.8–3.8\times$ higher.

%\clearpage

%\begin{figure*}[!t]
%\begin{center}
%\includegraphics[width=0.9\textwidth]{Figures/Ti_coexistence.pdf}
%\captionsetup{width=0.5\linewidth}
%\caption{\label{fig:Ti_coexistence} \textbf{Solid-liquid coexistence. A.} In the equilibrium view (no kinetic effects), when the Hugoniot intersects the the melt line, melt initiates. Increasing shock pressure accesses states along the melt line with mixed solid (S) and liquid (L) until full melt is achieved \cite{samanta2014}. \textbf{B.} Evolution of liquid (grey shaded) vs bcc diffraction (red) as a function of increasing shock pressure.
%}
%\end{center}
%
%\end{figure*}
\end{document}